\documentclass[usenatbib,a4paper]{mn2e}
\usepackage{natbib,psfig}

\catcode`\@=11
\def\gta{\ifmmode{\,\mathrel{\mathpalette\@versim>\,}}
    \else{$\,\mathrel{\mathpalette\@versim>}\,$}\fi}
\def\lta{\ifmmode{\,\mathrel{\mathpalette\@versim<\,}}
    \else{$\,\mathrel{\mathpalette\@versim<}\,$}\fi}
\def\@versim#1#2{\lower 2.9truept \vbox{\baselineskip 0pt \lineskip
    0.5truept \ialign{$\m@th#1\hfil##\hfil$\crcr#2\crcr\sim\crcr}}}
\catcode`\@=12  

\usepackage{graphics} \usepackage{amssymb}
\usepackage{setspace} 

\newcommand{\beq}{\begin{equation}}
\newcommand{\eeq}{\end{equation}}

\newcommand{\figref}[1]{Fig.~\ref{#1}}

\def\Gyr{\,{\rm Gyr}}

\def\vlos{v_\parallel}
\def\vperp{v_\perp}
\def\Flos{F_\parallel}
\def\d{{\rm d}}
\def\kpc{\,{\rm kpc}}
\def\kms{\,{\rm km\,s}^{-1}}
\def\deg{^\circ}

\title[Locating the orbits delineated by tidal streams]
{Locating the orbits delineated by tidal streams}

\author[A. Eyre and J. Binney]
{Andy Eyre and James Binney\\
Rudolf Peierls Centre for Theoretical Physics, Keble Road, Oxford OX1
3NP, UK}

\begin{document}

\date{Draft, May 23, 2008}

\pagerange{\pageref{firstpage}--\pageref{lastpage}} \pubyear{2008}

\maketitle

\label{firstpage}

\begin{abstract}
  We describe a technique that finds orbits through the Galaxy that
  are consistent with measurements of a tidal stream, taking into
  account the extent that tidal streams do not precisely delineate
  orbits. We show that if accurate line-of-sight velocities are
  measured along a well defined stream, the technique recovers the
  underlying orbit through the Galaxy and predicts the distances and
  proper motions along the stream to high precision. As the error bars
  on the location and velocities of the stream grow, the technique is
  able to find more and more orbits that are consistent with the data
  and the uncertainties in the predicted distances and proper motions
  increase. With radial-velocity data along a stream $\sim40^\circ$
  long and $\lta0.3^\circ$ wide on the sky accurate to $\sim1\kms$ the
  precisions of the distances and tangential velocities along the
  stream are 4 percent and $5\kms$, respectively. The technique can be
  used to diagnose the Galactic potential: if circular-speed curve is
  actually flat, both a Keplerian potential and $\Phi(r) \propto r$
  are readily excluded.  Given the correct radial density profile for
  the dark halo, the halo's mass can be determined to a precision of 5
  percent.
\end{abstract}


\section{Introduction}

Deep optical surveys of the Milky Way and other Local-Group galaxies have
uncovered numerous stellar streams
\citep{Odenkirchen,Majewski04,fostreams,Ibata}.
The Leiden-Argentine-Bonn survey of the Galaxy in the $21\,$cm line of
hydrogen \citep{LAB} contains many similar streams. In all probability both
stellar and gaseous streams have been tidally torn from orbiting bodies, and
as such delineate the orbits of those bodies around the Galaxy
\citep{JohnstonHB,Odenkirchen03,Choi}. Newton's laws of motion severely constrain
the readily observable quantities along an orbit in the sky, namely the
sequence of positions on the sky $[l(u),b(u)]$ and the corresponding
line-of-sight velocities, $v_\parallel(u)$, where $u$ is a parameter that
varies monotonically along the stream \citep[][hereafter Paper
I]{complexA,paper1}.  In fact, if the observables are known to reasonable
accuracy, data for a single stream can strongly constrain the Galaxy's
gravitational potential, and once the potential is known, the distance and
proper motion at each point on the stream can be predicted with an accuracy
that far exceeds anything likely to be possible by conventional astrometry
(Paper I).

From the work of Paper I it emerges that the major limitation on the
diagnostic power of streams is that streams do {\it not\/} precisely
delineate individual orbits \citep{Choi}. This paper is devoted to exploring
the extent to which this limitation can be overcome. In Section~\ref{sec:norbit}
we illustrate the extent of the problem, in
Section~\ref{sec:identify} we introduce significant improvements to the
methodology of Paper I and use these to identify orbits that are
consistent with a given body of data. In Section~\ref{sec:test} we test
this approach. In Section~\ref{sec:potential} we examine our ability to
correctly diagnose the Galactic potential. Section~\ref{sec:conclusions}
sums up and discusses directions for future work.

Except where stated otherwise, orbits and reconstructions are
calculated using the Galactic potential of Model II in \cite{gd2},
which is a slightly modified version of a halo-dominated potential
described by \cite{dehnenbinney}. We take the distance to the Galactic
centre to be $8\kpc$ and from \cite{ReidB} (for $V$ and $W$) and
\cite{Hipp} we take the velocity of the Sun in the Galactic rest frame
to be $(U,V,W)=(10.0,241.0,7.6)\kms$.

\begin{figure*}
\centerline{\psfig{file=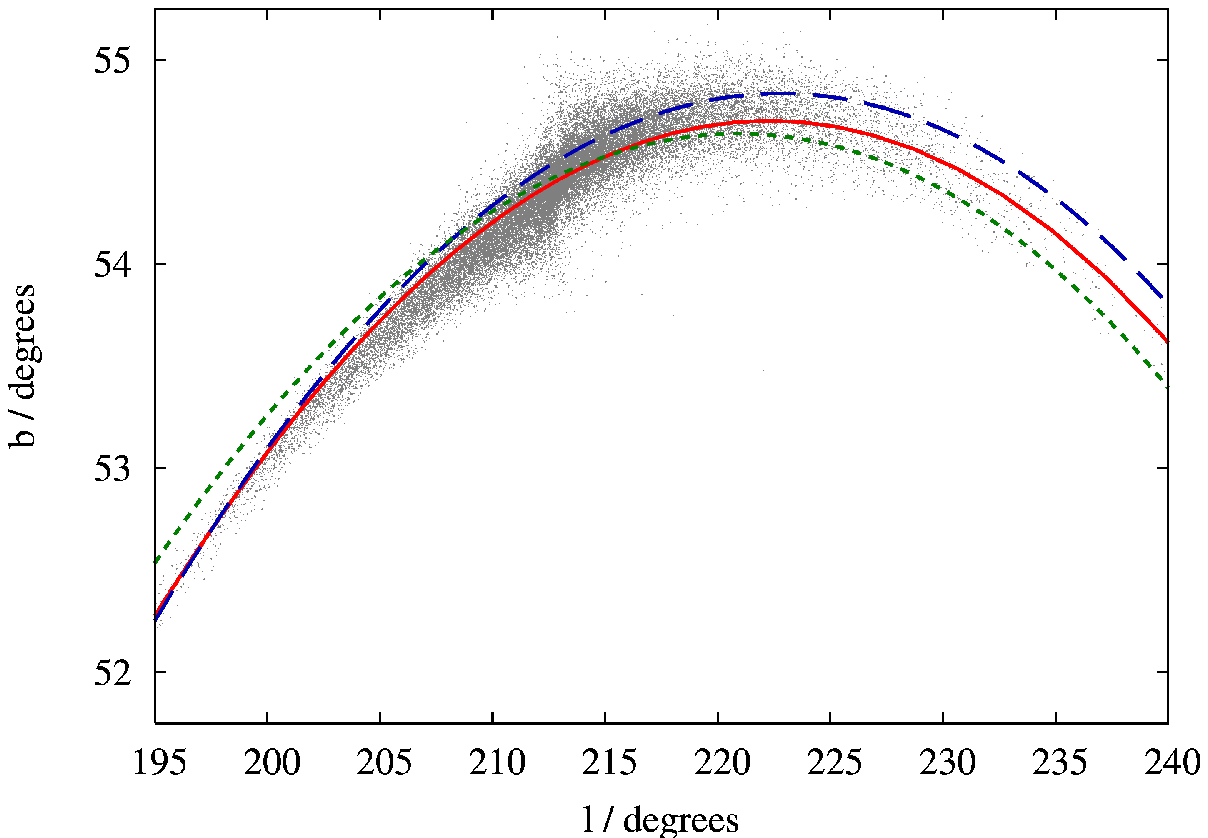,width=.48\hsize}\quad
\psfig{file=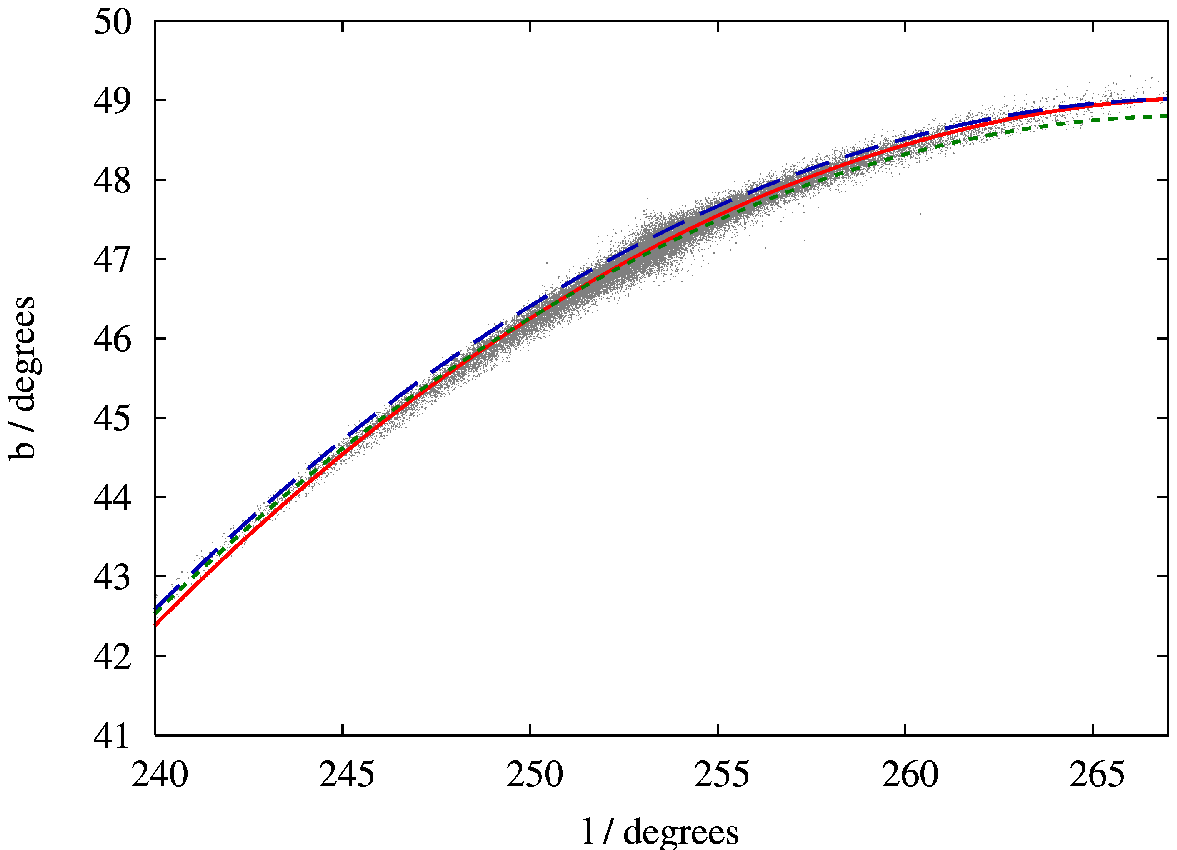,width=.48\hsize}}
\centerline{\psfig{file=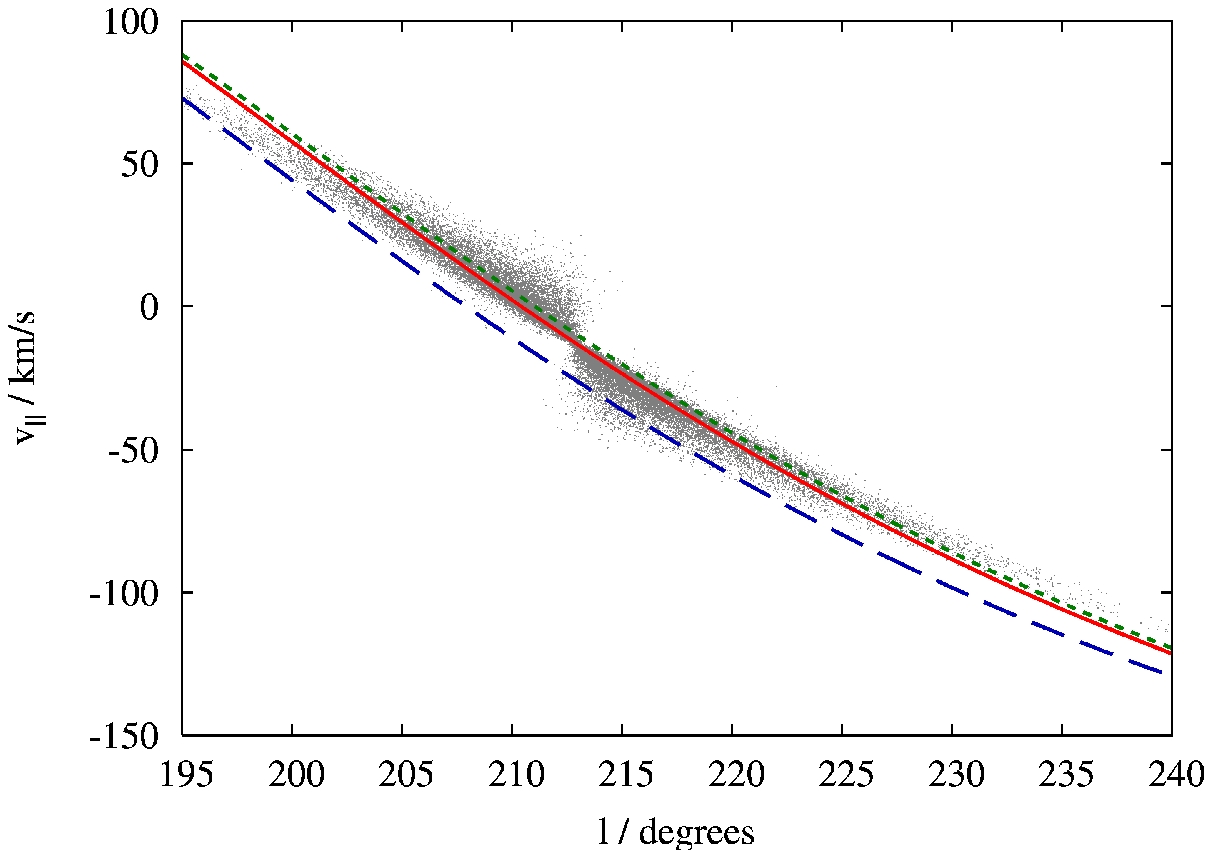,width=.48\hsize}\quad
\psfig{file=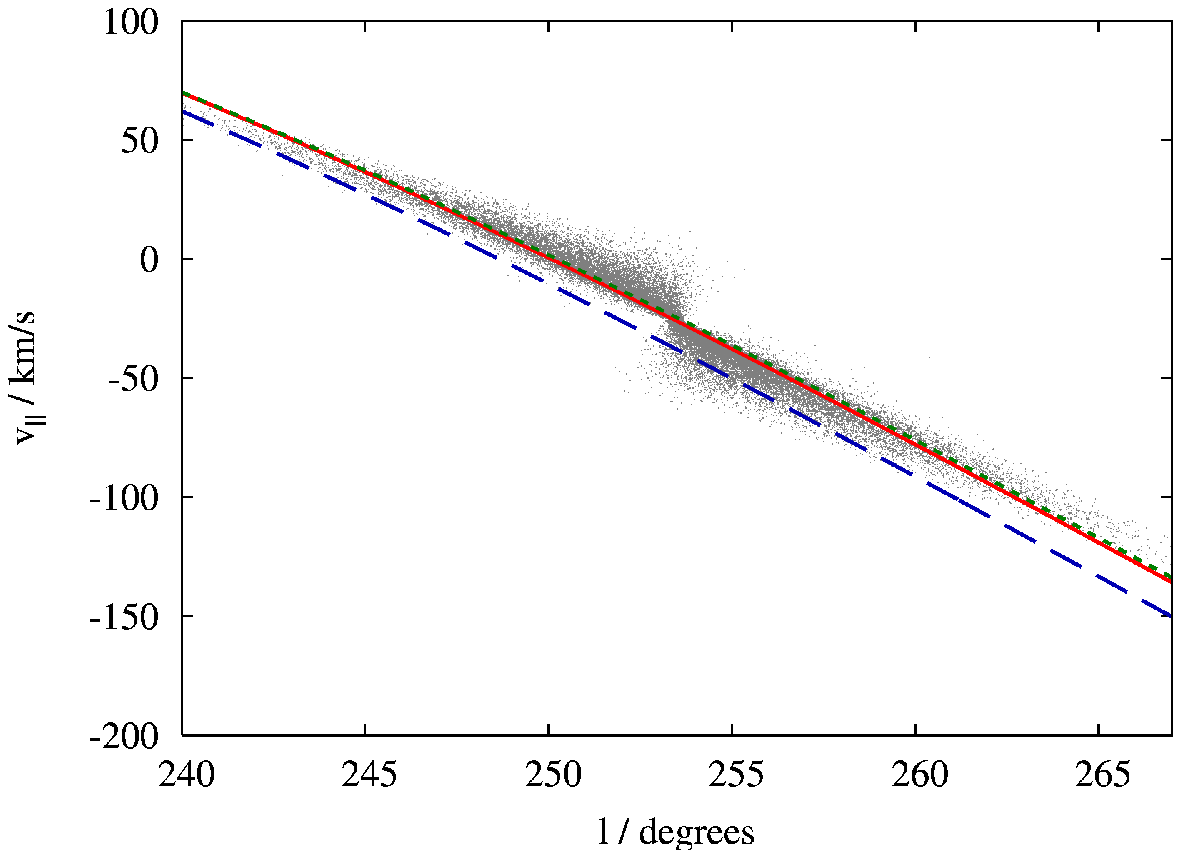,width=.48\hsize}}
\caption{Full (red) lines: the orbit of a progenitor of an Orphan-like
  stream.  Broken (green/blue) lines: orbits of a star now seen
  at either end of the tidal tail.  Points: particles tidally stripped
  from an N-body model of the Orphan-like progenitor.  Upper panels:
  distributions on the sky; lower panels: line-of-sight
  velocities. The N-body model had $60\,000$ particles set up as a
  King model with $W=2$, $r_0 = 13.66 \, {\rm pc}$ and $ M_0 = 9381
  M_{\sun}$, on the orbit detailed in Table~\ref{orbittab}, and
  evolved for $9.43\Gyr$. The particles were advanced in time by the
  ``FVFPS'' tree code of Londrillo et al. (2003).}
\label{nbody}
\end{figure*}

\section{The problem}\label{sec:norbit}

The full curves in \figref{nbody} show an orbit superficially similar
to that underlying the Orphan Stream \citep{orphan} from two viewing
locations -- the position of the Sun and a position $120^\circ$ further
round the solar circle. Also shown in each projection are the
locations of particles tidally stripped from a self-gravitating N-body
model of a cluster launched onto the given orbit. Clearly the particles
provide a very useful guide to the orbit of the cluster, but they do
not precisely delineate it.  Moreover, the relationship of the orbit
to the stream depends on viewing angle. The line-of-sight velocities
of stream particles have a similar relationship to the orbit's
line-of-sight velocity.  Hence even with perfectly error-free data,
the orbit we seek will not coincide with the stream, and before we can
fully exploit the dynamical potential of streams, we have to
understand how to infer the location of an underlying orbit from
measurements of the stream.

With what precision could the track of an orbit be specified from the
positions of the particles in \figref{nbody}? First we need to be clear that
{\it any\/} orbit will do. Generally it will be convenient to use the orbit
that passes through some  point that lies near the centre of the observed
stream, both on the sky and in line-of-sight velocity. In some circumstances
this orbit will closely approximate the orbit of the centre of mass of the
stream's progenitor, but there is no requirement that this is so. Since a
central point on the orbit is chosen at will, this point is associated with
vanishing error bars. As one moves away from this point, either up or down
the stream, it becomes more uncertain where the chosen orbit lies, and the
size of our error bars must increase. Hence the region of $(l,b,v_\parallel)$
space to which the orbit is confined by the observations is widest at its
extremities and shrinks, usually to a point, at its centre. We call this
the ``bow-tie region''.

In the top right panel of \figref{nbody} the leading and trailing streams are
not offset for most of the span so we may assume that the orbit through the
stream's centre runs right down the centre of the stream.  In the lower
panels the stream has a kink at the progenitor and our best guess is that the
orbit through the point where the two halves of the tail touch runs near the
lower edge of the left-hand half and near the upper edge of the right-hand
half. In every case the error bars on the location of the orbit grow from
zero at the centre to roughly half width of the stream at
its ends. Quantitatively, the largest  error is then $0.15\deg$ for
the top-left panel, $\sim0.25\deg$ for the top-right panel, and  $\sim5\kms$
in the bottom panels.

\section{Identifying dynamical orbits}\label{sec:identify}

Paper I showed that given an orbit's projection onto the sky
$[l(u),b(u)]$ and the corresponding line-of-sight velocities
$v_\parallel(u)$, the remaining phase space coordinates can be recovered by
solving the differential equations
 \begin{eqnarray}\label{eq:basicde}
{\d t\over\d u}&=&{1\over2\Flos}\left({\d\vlos\over\d u}-
\sqrt{\left({\d\vlos\over\d u}\right)^2-4s\Flos}\,\right)\nonumber\\
{\d s\over\d u}&=&\vlos{\d t\over\d u}.
\end{eqnarray}
 Here $u$ is distance on the sky down the
projected orbit, $s(u)$ is three-dimensional distance to the orbit, 
$t(u)$ is the time at which the orbiting body reaches the given point on the
orbit and $F_\parallel$ is the component of the Galaxy's gravitational field
along the line of sight. The reference frame used is the inertial frame in
which the Galactic centre is at rest; consequently the velocities
$v_\parallel$ are obtained by subtracting the projection of the Sun's motion
along the given line of sight from the measured heliocentric velocities.

If the input data used to solve equations (\ref{eq:basicde}) are not
derived from an orbit in the same force field as is used to derive
$F_\parallel$, the reconstructed phase-space coordinates will not
satisfy the equations of motion. Paper I observed that violation of
the equations of motions might cause the reconstructed solution to
violate energy conservation, and therefore used rms energy
variation down the track as a diagnostic for the quality of a
solution. However, energy conservation is necessary but not sufficient
to qualify a track as being an orbit. Here we construct a diagnostic
quantity from residual errors in the equations of motion themselves,
since orbits are defined to be solutions of these equations.

We first derive the equations of motion. In the coordinates $(s, b, l)$, the
canonically  conjugate momenta are
 \begin{eqnarray}\label{eq:momenta}
p_s &=& \dot{s},\nonumber\\
p_b &=& s^2 \dot{b},\\
p_l &=& s^2 \cos^2b \, \dot{l}.\nonumber
\end{eqnarray}
The Hamiltonian is therefore
\beq
{\rm H} = \frac{1}{2}p_s^2 + \frac{1}{2}\frac{p_b^2}{s^2}
+ \frac{1}{2}\frac{p_l^2}{s^2 \cos^2 b} + \Phi(s, b, l),
\eeq
 and the equations of motion are
\begin{eqnarray}\label{eq:motion}
\dot{p_s} &=& \ddot{s} = \frac{p_b^2}{s^3} + \frac{p_l^2}{s^3 \cos^2 b} - 
\frac{\partial \Phi}{\partial s},\nonumber\\
\dot{p_b} &=& \ddot{b}s^2 + 2 \dot{b} \dot{s} s =
- \frac{p_l^2}{s^2}\frac{\sin b}{\cos^3 b} - \frac{\partial \Phi}{\partial b},\\
\dot{p_l} &=& s^2 \cos^2 b \ddot{l} - 2 s^2 \dot{l} \dot {b} \sin b \cos b +
2 s \dot{s} \dot{l} \cos^2 b = -\frac{\partial\Phi}{\partial l}.\nonumber
\end{eqnarray}

As in Paper I, when solving equations (\ref{eq:basicde}) extensive use is
made of cubic-spline fits to the data. In the examples presented in Paper I
natural splines were used in order to avoid specifying the gradient of the
data at its end points. Significantly improved numerical accuracy can be
achieved by taking the trouble to specify these gradients explicitly.  Given
input data, we estimate the quantity ${\d l}/{\d b}$ at the end points by fitting
a quadratic curve through the first three and last three points.
${\d l}/{\d b}$ is then computed at the location of the middle point of each
set, and the very first and very last points are considered `used' and thrown
away. This quantity is then used in the geometric relation
 \beq
 \frac{\d u}{\d b}= \pm \sqrt{ 1 + \cos^2 b \left(\frac{\d l}{\d b}\right)^2},
 \eeq
 to compute
${\d b}/{\d u}$ at the ends of the track. The sign ambiguity is resolved by
inspection of the directionality of the input data. We then use the geometric
relation
 \beq
 \frac{\d l}{\d u} = \pm \sec b \sqrt{ 1 -
  \left(\frac{\d b}{\d u}\right)^2},
\eeq
 to obtain ${\d l}/{\d u}$ at the ends of the track, where the sign ambiguity is
resolved in the same way. We are now able to fit cubic splines through the
input tracks, with the slopes at the end of the $l(u)$ and $b(u)$ tracks
given as above, but at this stage the track of $v_\parallel(u)$ is fitted
with a natural spline. The reconstruction equations (\ref{eq:basicde}) are
now solved for $t(u)$, which is then fitted by a cubic spline, with the
slopes at the ends given explicitly by (\ref{eq:basicde}).

We can now compute $l(t)$ and  $b(t)$ and fit splines to them, with the
slopes at the ends computed from ${\d l}/{\d u}$ and ${\d b}/{\d u}$
by the chain rule. The momenta (\ref{eq:momenta}) are now calculated
explicitly, using the derivatives of the splines $l(t), b(t)$ in place
of $\dot{l}, \dot{b}$. The slopes at the endpoints,
${\d v_\parallel}/{\d u}$, can now be calculated from (\ref{eq:motion})
and ${\d t}/{\d u}$; the $v_\parallel(u)$ spline is refitted
using these boundary conditions, the reconstruction repeated, and
the momenta recalculated.

The left- and right-hand sides of the equations of motion (\ref{eq:motion})
are calculated explicitly. For each equation of motion we define a residual
 \beq
R(t) = \dot{p}_{\rm lhs}(t) - \dot{p}_{\rm rhs}(t).
\eeq
These residuals are used to compute, for each equation of motion,
 the diagnostic quantity
 \beq
D = \log_{10} \left(\frac{\int_{t_1}^{t_2}\d t\, R(t)^2}
{\int_{t_1}^{t_2}\d t\, \dot{p}_{\rm lhs}^2}\right), \label{eq:diag}
\eeq
 where the residuals have been normalised by the mean-square acceleration and
the times
$t_1$ and $t_2$  correspond to the fifth and
fifth-from-last input data points; the residual errors from the end regions,
$0 < t < t_1$ and $t_2 < t < t_{\rm max}$, tend to dominate the integrated
quantity and are not easily reduced by modifying the input; they are
therefore excluded.  The largest of the three values for $D$ is used as
the diagnostic quantity for that particular input.

\subsection{Parametrising tracks}

Our strategy for identifying a stream's underlying orbit is to compute the
diagnostic $D$ (eq.~\ref{eq:diag}) for a large number of candidate tracks,
and to find which candidates yield values of $D$ consistent with their being
dynamical orbits.

We start by specifying a baseline track across the sky $[l_{\rm b}(u'),b_{\rm
b}(u')]$, where $u'$ is a parameter that increases monotonically down the
track from $-1$ to $1$.  Similarly, we specify associated baseline line-of-sight velocities
$v_{\parallel\rm b}(u')$. The baseline track is required to pass through
the error bars of every data point. 

All candidate tracks should be smooth because orbits are. We satisfy this
condition by expressing the difference between the baseline track and a
candidate track as a low-order polynomial in $u'$.
For  streams that cover a wide range of longitudes, 
the parametrisation of candidate tracks is achieved
by slightly changing the values of $b$ and $v_\parallel$ associated with a
given value of $l$ from the values specified by the  baseline functions.
That is we write
 \begin{eqnarray}\label{eq:paramtrack}
b(u')& =& b_{\rm b}(u') + \sum_{n = 0}^N b_n T_n(u'),\nonumber\\
v_\parallel(u')& =& v_{\parallel\rm b}(u') + \sum_{n = 0}^N a_n T_n(u'),
\end{eqnarray}
 where $T_n$ is the $n^{\rm{th}}$-order Chebyshev polynomial of the first
kind and $a_n$ and $b_n$ are free parameters.  These $2N$ parameters are
coordinates for the space of tracks that we have to search for orbits. When a
stream does not stray far from the Galactic plane, candidate tracks are best
parametrised by adjusting the baseline values of $b$ and $v_\parallel$ at
given longitude.  In all examples in this paper, the series in equations
(\ref{eq:paramtrack}) are truncated after $N=10$. A larger number of terms
allows the correction function to produce tracks that represent orbits
better, but makes the search procedure computationally more expensive.  The
number of terms used is a compromise between these considerations.

The space of tracks is defined by the $a_n$ and $b_n$ and
one extra parameter, the distance to the stream, $s_0$, at the
starting point $u=0$ for the integration of equations
(\ref{eq:basicde}).

We shall henceforth denote a point in the ($2N+1$)-dimensional space of
parameters by $\chi$.  Each $\chi$ is associated with a complete
specification of all six phase-space coordinates for every point on the
candidate orbit: $l$, $b$ and $v_\parallel$ follow from the parametrisation and the
remaining coordinates are obtained by solution of the differential equations
(\ref{eq:basicde}). Consequently, each $\chi$ corresponds to a value of
$D$ (eq.~\ref{eq:diag}) that quantifies the extent to which the
phase-space coordinates deviate from a dynamical orbit in the given potential.

\subsection{Searching parameter space}

Dynamical orbits are found by minimising the sum 
\beq\label{eq:Dtot}
D' (\chi)=D(\chi)+p(\chi)
\eeq
 where $p(\chi)$ is the sum of the penalty functions:
\beq
p(\chi) = \sum_i p_{i,\rm pos} + \sum_i p_{i,\rm vel} + p_{s},
\label{penaltyfn}
\eeq
where
\beq\label{eq:ppos}
p_{i,\rm pos} =\cases{\Delta_{i,\rm pos}&if $\Delta_{i,\rm pos}>1$\cr
0&otherwise}
\eeq
 with
 \beq
\Delta_{i,\rm pos}={\left| b(l_i) - b_{\rm b}(l_i) \right|\over\delta b
(l_i)}.
\eeq
 Here $\delta b_i$ is the width in $b$ of the bow-tie region at $l_i$. Similarly
\beq\label{eq:pvel}
p_{i,\rm vel}(l) =\cases{\Delta_{i,\rm vel}&if $\Delta_{i,\rm vel}>1$\cr
0&otherwise}
\eeq
 with 
\beq
\Delta_{i,\rm vel}={\left| v_\parallel(l_i) - v_{\parallel\rm b}(l_i) \right|
\over\delta v_\parallel (l_i)}.
\eeq
 Prior information about the
distance to the stream is used by specifying the penalty function
$p_{\rm s}$ to be
\beq\label{eq:defsps}
p_{\rm s} =
\left\{
\begin{array}{ll}
\beta \left| s_0 - s_{0\rm b}\right|/\delta s &
\quad\mbox{if $\left| s_0 - s_{0\rm b} \right| > \delta s$} \\
0 &\quad\mbox{otherwise},
\end{array} \right. 
\eeq
 where $\delta s$ is the half-width of the allowed range in the distance
$s_0$ to the starting point of the integrations and $s_{0\rm b}$ is the
baseline value of $s_0$. These definitions are such that $p(\chi)=0$ so long
as the track lies within the region that is expected to contain the orbit,
and rises to unity, or in the case of $p_{\rm s}$ to $\beta$, on the boundary
and then increases continuously as the orbit leaves the expected region.

In practical cases the prior uncertainty in distance is large, and the
obvious way to search for orbits is to set $\delta s$ to the large value that
reflects this uncertainty and then set the algorithm described below to work.
It will find candidate orbits for certain distances. However, we shall see
below that it is more instructive to search the range of possible distances
by setting $\delta s$ to a small value such as $0.5\kpc$ and searching for
orbits at  each of a grid of values of $s_{0\rm b}$. In this
way we not only find possible orbits, but we show that no acceptable orbits
exist outside a certain range of distances. In  this procedure the  logic
underlying $\delta s$ is very different from that underlying $\delta b$ and
$\delta v_\parallel$.

Since $p(\chi)$ is added to the {\it logarithm\/} of the rms errors in the equations
of motion and increases by of order unity at the edge of the bow-tie region,
the algorithm effectively confines its search to the bow-tie region, where
$p=0$. Thus at this stage we do not discriminate against orbits that graze
the edge of the bow-tie region in favour of ones that run along its centre.
Our focus at this stage is on determining for which distances dynamical
orbits can be constructed that are compatible with the data.  Once this has
been established, distances that lie outside some range can be excluded from
further consideration.

The space of candidate tracks $\chi$ is 21-dimensional, so an
exhaustive search for minima of $D'$ (eq.~\ref{eq:Dtot}) is
impractical. Furthermore, the landscape specified by $D'$ is
complex. Some of this complexity is physical; the space should contain
continua of related orbits, and ideally $D' \to-\infty$ at
orbits. Hence deep trenches should criss-cross the space.
Superimposed on this physical complexity is a level of numerical noise
arising from numerical limitations in the computation of $D'
(\chi)$. The limitations include the use of finite step sizes in the
solution of equations (\ref{eq:basicde}) and the subsequent evaluation
of $D' (\chi)$, as well as the difficulty in representing a true
orbital track with a collection of sparse input points interpolated
with splines. In practice numerical noise sets a lower limit on the
returned values of $D' (\chi)$.

On account of the complexity of the landscape that $D' $ defines, ``greedy''
optimisation methods, which typically follow the path of steepest descent,
are not effective in locating minima. The task effectively becomes one of
global minimisation, which is a well studied problem in optimisation.

We have used the variant of the Metropolis ``simulated annealing'' algorithm
described in \cite{Press}, which uses a modified form of the downhill simplex
algorithm.  In the standard simplex algorithm, the mean of the values of the
objective function over the vertices decreases every time the simplex
deforms. In the Press et al.\ algorithm the simplex has a non-vanishing
probability of deforming to a configuration in which this mean is higher than
before.  Consequently, the simplex has a chance of crawling uphill out of a
local minimum. The probability that the simplex crawls uphill is controlled
by a ``temperature'' variable $T$: when $T$ is large, uphill moves are likely,
and they become vanishingly rare as $T\to0$. During annealing the value of
$T$ is gradually lowered from an initially high value towards zero.

One vertex of the initial simplex is some point $\chi_{\rm guess}$ and
the remaining $2N+1$ vertices are obtained by incrementing each
coordinate of $\chi_{\rm guess}$ in turn by a small amount. For the
coefficients of $T_0$ this increment is approximately the size of the
allowed half widths, $\delta s, \delta v$ and $\delta b$. Increments
for coordinates representing coefficients of higher-order $T_n$ are
scaled as $1/n$. The overall size of these increments is therefore set by
the size of the region within which we believe the global minimum to
lie. It is important to note that in each generation of a simplex, the
increments should independently have equal chance of being added to or
subtracted from the values of $\chi_{\rm guess}$ so that no part of the
parameter space is unfairly undersampled. The algorithm makes
tens of thousands of deformations of the simplex while $T$ is linearly
reduced to zero.

This entire process is repeated some tens of times, after which we have a
sample of local minima that are all obtained from $\chi_{\rm guess}$.

We now update $\chi_{\rm guess}$ to the location of the lowest of the
minima just found and initiate a new search.  The entire process is
repeated until the value of the diagnostic function $D' (\chi)$ hits a
floor. When this floor lies higher than the numerical-noise floor, the
attempt to find an orbit that is consistent with the assumed inputs
has been a failure and we infer that no such orbit exists. When the
floor coincides with the numerical-noise floor, we conclude that the
corresponding $\chi$ specifies an orbit that is compatible with the
inputs. An approximate value for the numerical noise floor for a given
problem may be obtained as follows: given input that perfectly
delineates an orbit in the potential in use, the value of $D'$
returned at the correct distance is approximately the numerical-noise
floor. Conclusive proof that a candidate track with a particular value of
$D'$ is an orbit can be obtained by integrating the equations of
motion from the position and velocity of any  point on the
track and ensuring that the time integration essentially recovers the track.

On account of the stochastic nature of the algorithm, an attempt to
find a solution at a particular distance occasionally sticks at a
higher value of $D'$ than the underlying problem allows. This
condition is identified by scatter in the values of $D'$ reached on
successive attempts and by inconsistency of these values with the
values of $D'$ achieved for nearby distances---we see from
\figref{pd1} that the function underlying the minima is smooth. When
the magnitude of this scatter is significant, one can only confidently
declare an attempt to find an orbit a failure if the $D'$
achieved is consistently higher than the noise floor by more than the
scatter; since the diagnostic measure $D'$ quantifies the extent to
which a candidate track satisfies the equations of motion, by
definition, tracks with higher $D'$ than the noise floor plus scatter
cannot represent orbits.

When the observational constraints are weak, we expect several orbits to be
compatible with them. In particular, we will be able to find acceptable
orbits for a range of initial distances $s_0$.  It is therefore important,
for any given input, to run the algorithm starting from many different
values of $s_{0\rm b}$ with $\delta s$ set to prevent the algorithm straying
far from the specified $s_{0\rm b}$. In this way, the full range of allowable
distances can be mapped out, and dynamical orbits found for each distance in
that range. In the case of significant scatter about the noise-floor, the
range of distances at which valid orbits are found is the range
within which solutions yield values of $D'$ smaller than the noise-floor plus scatter.

Similar degeneracies in the parameters controlling the astrometry and
line-of-sight velocities are less of a  concern because if we have
orbits that differ in these observables, we simply concentrate on the orbit
that lies closest to the baseline track.

\begin{figure}
\centerline{\psfig{file=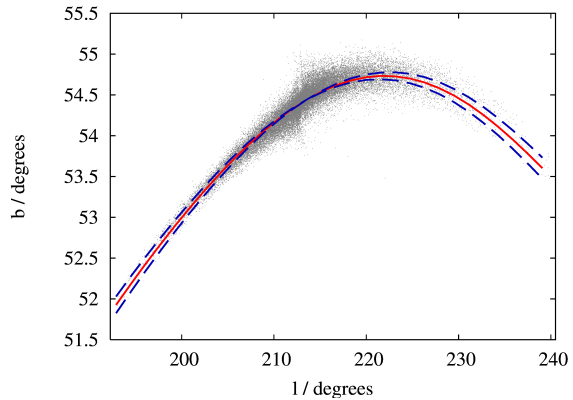,width=.93\hsize}}
\caption{Lines: Input track $b(l)$, as used for data sets PD2--PD6, along with
upper and lower bounds as set by the penalty function $p_{\rm pos}$ (eq \ref{eq:ppos}).
Dots: the projection of the N-body simulation onto the sky, from which the input
was derived.}
\label{sky-input}
\end{figure}

\begin{figure}
\centerline{\psfig{file=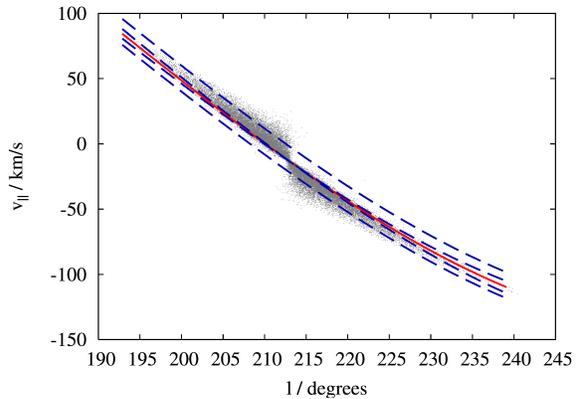,width=.93\hsize}}
\caption{Lines: Input track $v_\parallel(l)$, as used for data sets
  PD2--PD4, along with examples of upper and lower bounds as set for
  PD2 (narrow) and PD4 (wide). The bounds are enforced by the penalty function $p_{\rm vel}$
  (eq \ref{eq:pvel}). Dots: the projection of the N-body
  simulation onto the sky, from which the input was derived.}
\label{vr-input}
\end{figure}

\begin{figure}
\centerline{\psfig{file=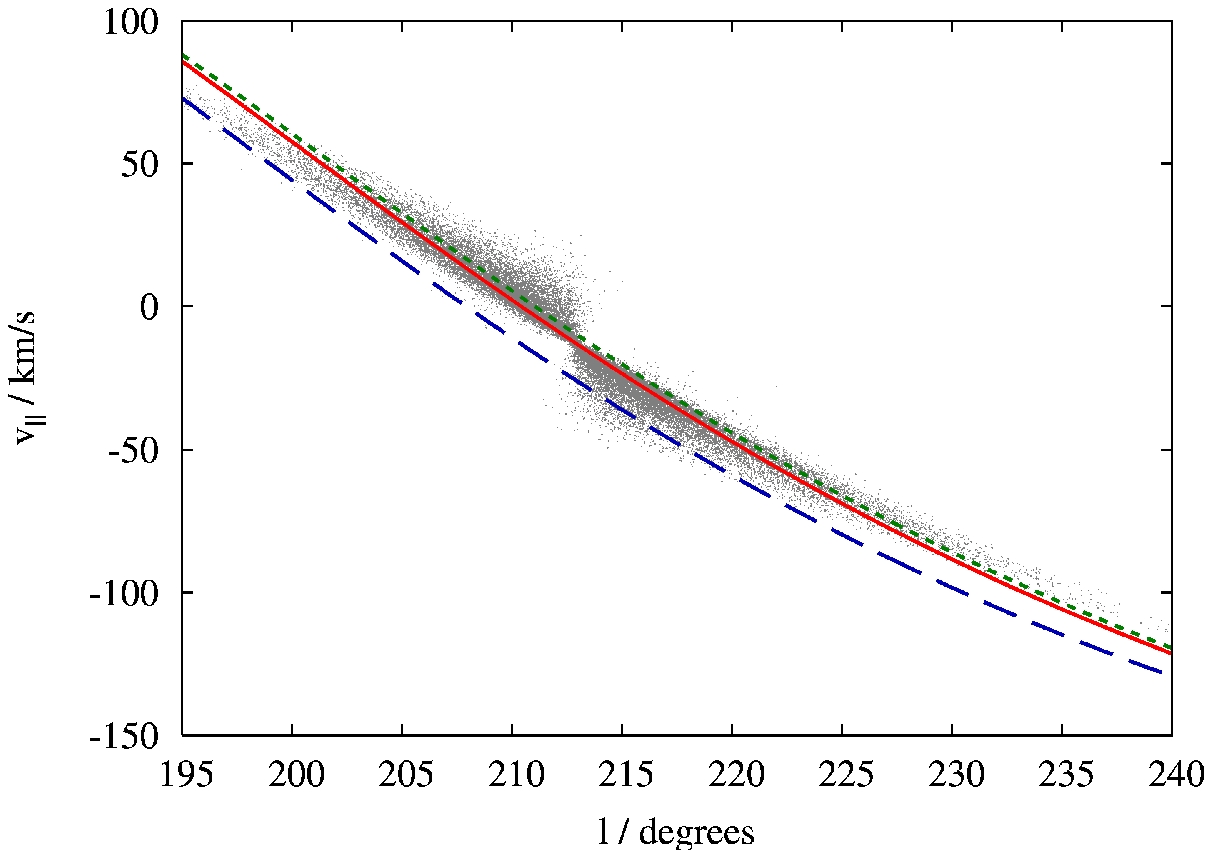,width=.93\hsize}}
\caption{Lines: Input track $v_\parallel(l)$, as used for data sets
  PD5 (dotted green) and PD6 (dashed blue). The unmodified input
  of PD2 (full red) is shown for comparison. Dots: the projection of the N-body
  simulation onto the sky, from which the input was derived.}
\label{vr-wrong}
\end{figure}

\section{Testing the method}\label{sec:test}

To test this method, we used the N-body approximation to the Orphan
Stream described in \figref{nbody} as our raw data. Sets of points of
$[l, b]$ and $[l, v_\parallel]$ were selected by eye to lie down the
middle of the stream. These sets were each fitted with a low-order
polynomial to ensure smoothness, and these polynomials were sampled at
30 points to produce the baseline input data $b(l)$ and
$v_\parallel(l)$.  To each data set we attached error estimates
$\delta b$ and $\delta v_\parallel$, which through the penalty
functions $p_{\rm pos}$ and $p_{\rm vel}$ (eqs \ref{eq:ppos} and
\ref{eq:pvel}) constrain the tracks that the Metropolis algorithm can
try.  Details of the resulting pseudo-data sets are given below, and
are summarised in Table~\ref{testtab}.

In one case, PD1, the above baseline input data were replaced by those
of a perfect orbit and $\delta b$ and $\delta v_\parallel$ set very
narrow ($6\,$arcsec and $2\times10^{-3}\kms$) in order to validate the
reconstruction algorithm.

\begin{figure}
\centerline{\psfig{file=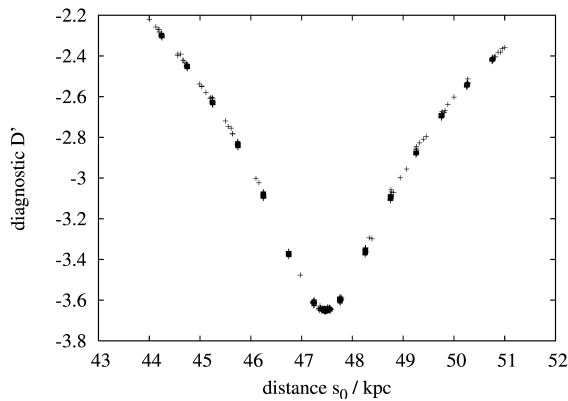,width=.93\hsize}}
\caption{Values of the diagnostic function $D'$
for candidate orbits reconstructed from the pseudo-data set PD1. Each group of crosses
is associated with one of 15 ranges within  which the starting distance $s_0$ was
constrained to lie. For each such range the candidate orbit was reconstructed from 280
trial tracks, and each track yields a cross. In this figure the crosses are
largely superimposed because the errors in the data set are small and there
is little scope for tweaking the track.}
\label{pd1}
\end{figure}

\begin{figure}
\centerline{\psfig{file=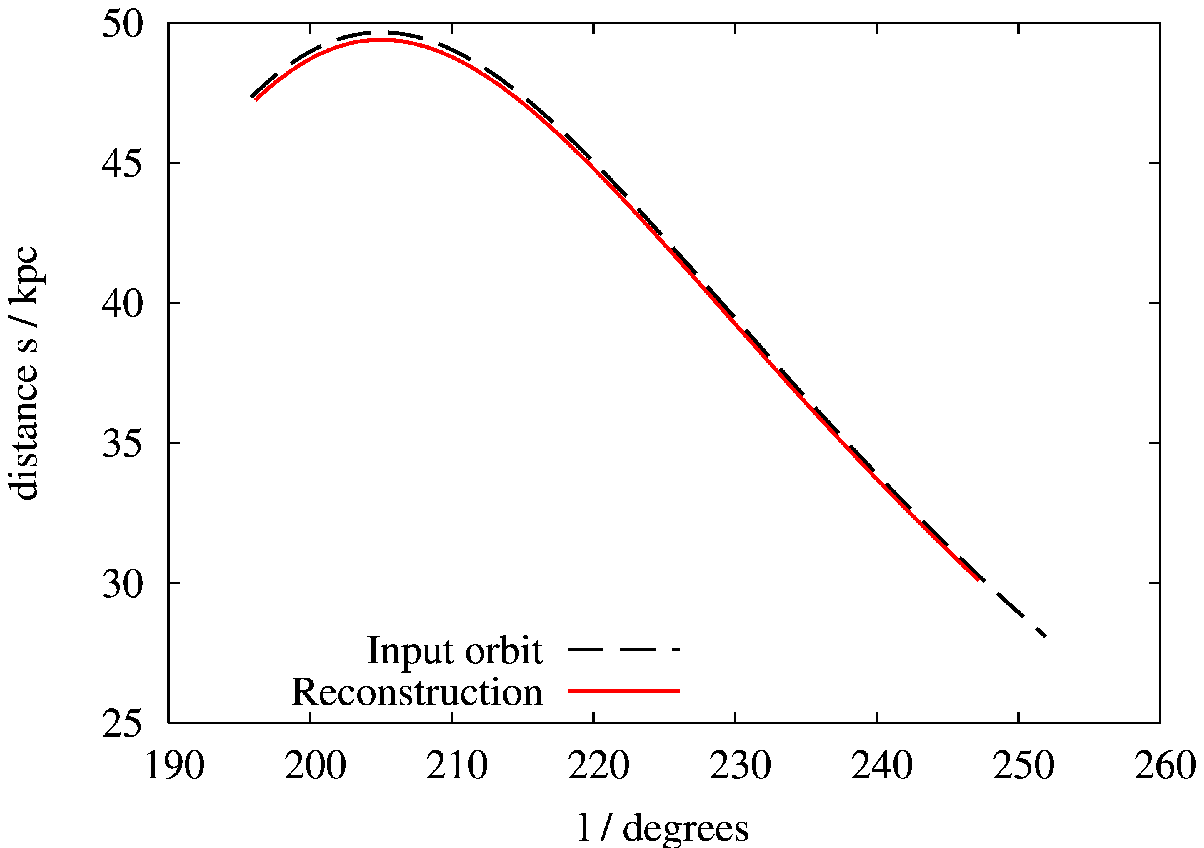,width=.93\hsize}}
\centerline{\psfig{file=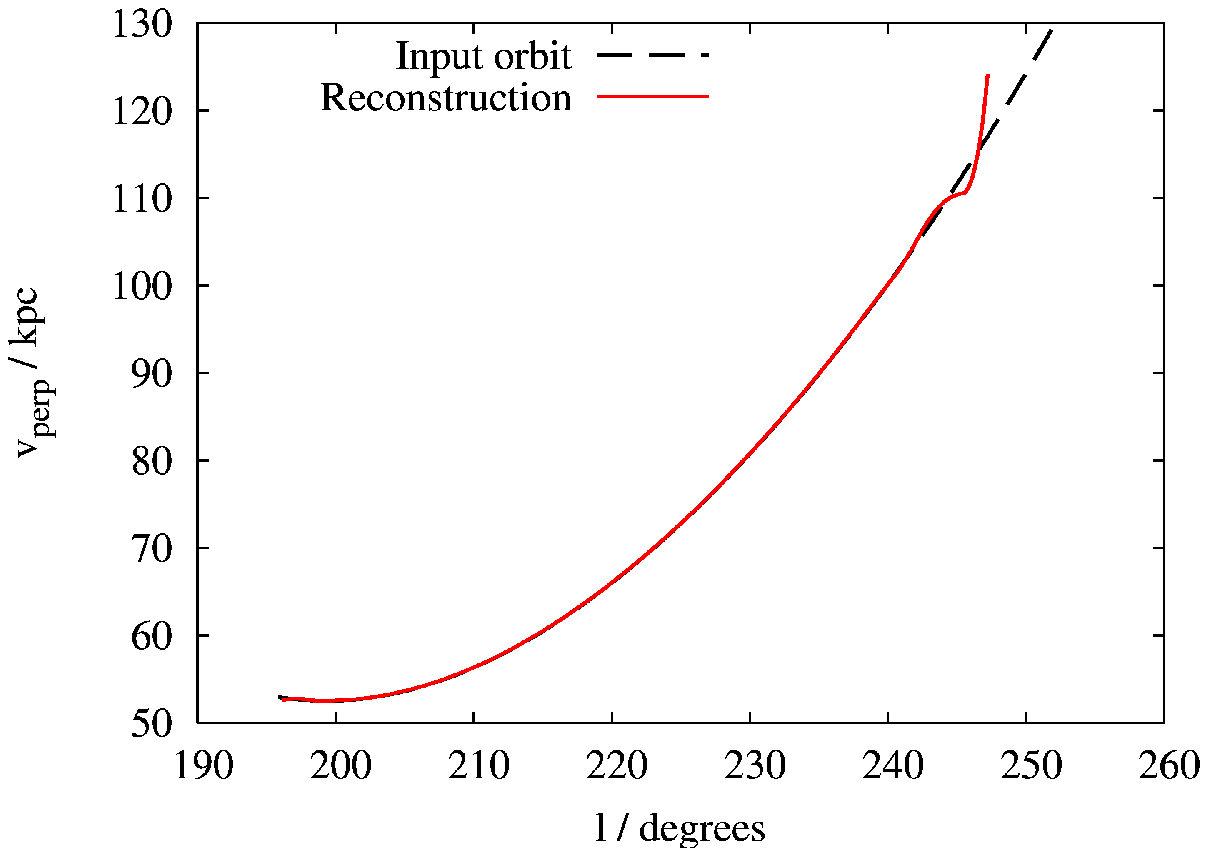,width=.93\hsize}}
\caption{The upper panel shows heliocentric distance, $s$,
versus galactic longitude, $l$, for the best reconstruction from \figref{pd1}
and for the true orbit from which the input was generated. The two curves
are close to overlying. The lower panel shows tangential velocity, $\vperp$,
for the best reconstruction and for the true orbit: the 
departure of the reconstruction from the orbit near the endpoints
is symptomatic of the problems near the endpoints that necessitate their
excision from the diagnostic.} 
\label{pd1-vs-truth}
\end{figure}

The uncertainty $\delta b(l)$ takes the same value for all the remaining
pseudo-data sets because we assume that the astrometry is sufficiently
precise for the error in position to be dominated by the offset of  the
stream from an orbit.  In all cases, $\delta b$ has a maximum value of
$0.15\deg$ at the ends of the stream, falling linearly to zero at the
position of the progenitor, consistent with the orbit of the
progenitor seen in \figref{nbody}. \figref{sky-input} shows this input
alongside the N-body data from which it was derived.

For the pseudo-data sets PD2 and PD3, $\delta v_\parallel$ is set to a maximum at
the ends of the stream, and falls linearly to a minimum of zero and
2km/s respectively, at the position of the progenitor. These examples
represent the case in which the uncertainty in radial velocity is
dominated by the orbital offset, and the case in which there is a
significant contribution of random error at a level that is easily
obtainable with a spectrograph. In one case, PD4,
$\delta v_\parallel$ is held fixed at $10\kms$ across the range. This
case is representative  of the imprecise radial-velocity information
currently available from the SDSS for stars in distant streams.
\figref{vr-input} shows the input for these data sets.

For the pseudo-data sets PD5 and PD6, we added to the baseline data systematic offsets in
$v_\parallel(l)$ to mimic systematic errors in radial velocity. $\delta v_\parallel$ varies between a maximum and a minimum as in PD2 and PD3,
with the values set to encompass the (assumed known) systematic bias.
\figref{vr-wrong} shows the input for these data sets.

The pseudo-data set PD7 is identical to that of PD2, except that the
number of raw $[l, v_\parallel]$ points was reduced to just three: one
at either end of the N-body stream, and one at the location of the
progenitor. A quadratic curve was perfectly fitted through these three
points, and sampled at 30 locations to produce the baseline
$v_\parallel(l)$ input. $\delta v_\parallel$ is set to the same
maximum value as PD2 at the outermost points; $\delta
v_\parallel$ is set to zero at the centre point. Only these three points
are allowed to contribute to the penalty function (\ref{eq:pvel}) in this
pseudo-data set.

\begin{figure}
\centerline{\psfig{file=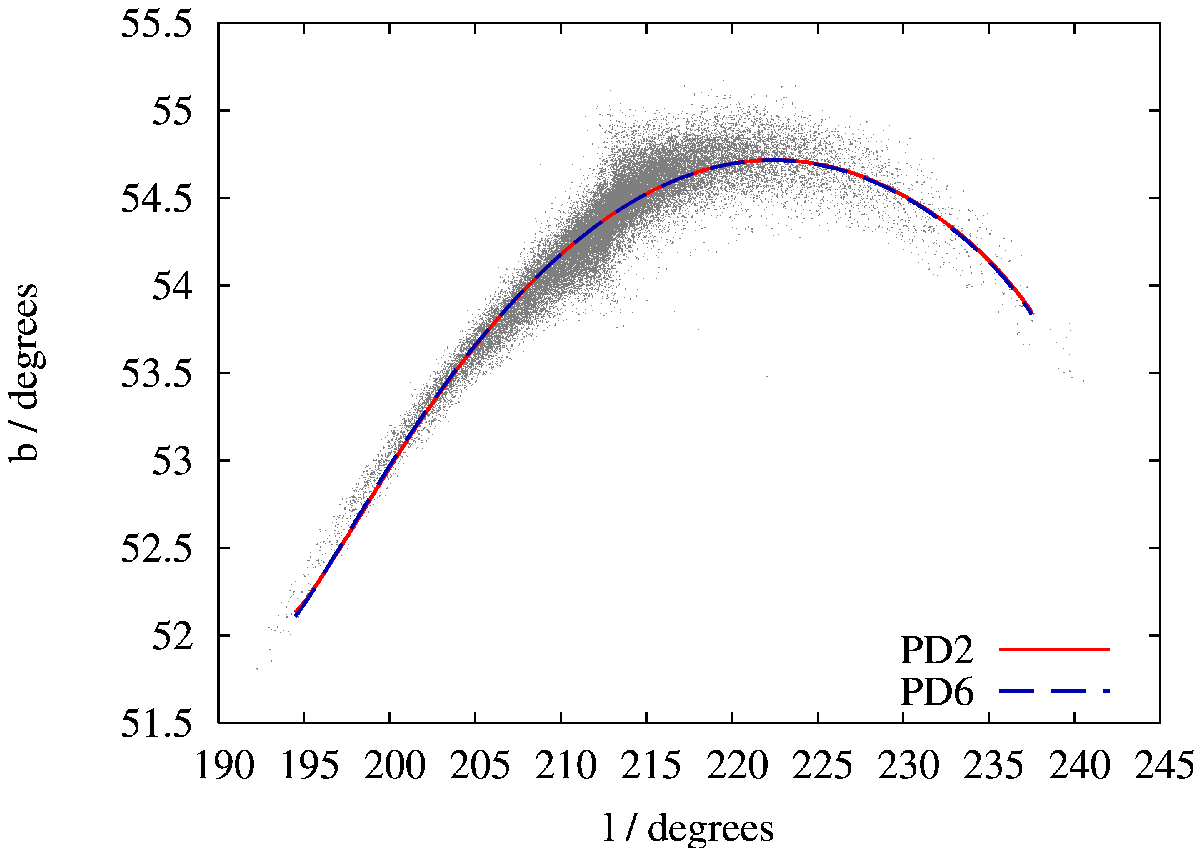,width=.93\hsize}}
\centerline{\psfig{file=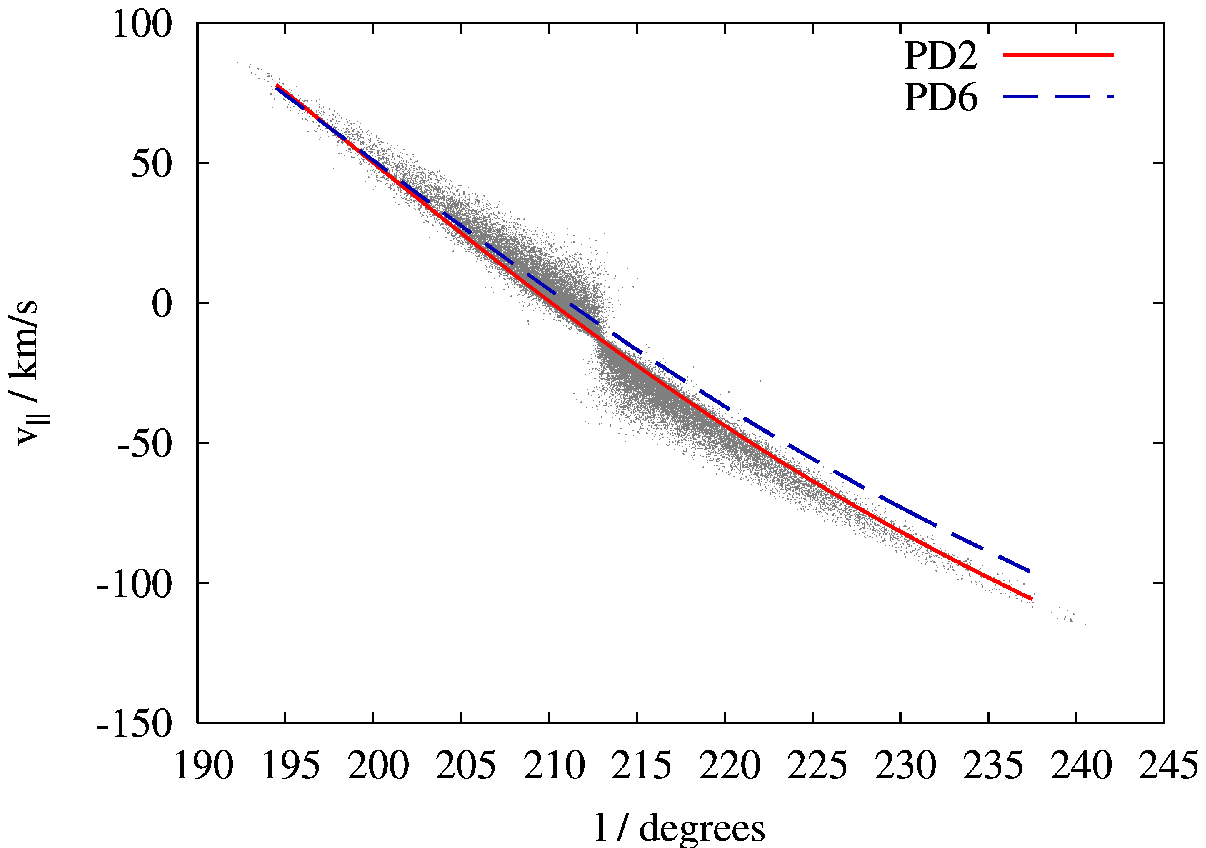,width=.93\hsize}} 
 \caption{Upper panel: projections onto the sky of two candidate orbits and
the N-body data from which the input was derived. The full curves show the
candidate orbit at $46\kpc$ from PD2, and the dashed curves show the
candidate orbit at $43\kpc$ from PD6,.  Lower panel: radial velocity down the
track for the same candidate orbits. The penalty function (eq.
\ref{penaltyfn}) forces the sky projection and radial velocity curves of
candidate orbits to be consistent with the data; the equivalent plots for all
other tracks are very similar to these examples.  }
\label{fig:sky-vr-examples}
\end{figure}

In all of the examples, the penalty function (eq. \ref{penaltyfn}) acts
to constrain the candidate tracks to be consistent with the data. The
contribution of the penalty function to $D'$ is therefore zero in all examples.
All candidate tracks are guaranteed to be consistent with the data, even
if they do not represent dynamical orbits. \figref{fig:sky-vr-examples} provides
example plots of $(l,b)$ and $(l,v_r)$ for candidate tracks from PD2 and
PD6 along with the raw N-body data from which the baseline data are derived.
Equivalent plots for all candidate tracks are similar to these.

\begin{table}
 \centering
 \begin{minipage}{85mm}
   \caption{Parameters of highlighted orbits from this paper. The
     coordinate system used is right-handed with $\hat{x}$ pointing
     away from the Galactic centre and $\hat{y}$ opposite the sense of
     Galactic rotation.}
  \label{orbittab}
  \begin{tabular}{@{}lccc@{}}
  \hline
  Orbit& position (x,y,z) $\rm / kpc$&
  velocity (x,y,z) $/\!\kms$\\
 \hline
 N-body Orphan & (28.1, -10.0, 34.0) & (-0.431, -0.179, -0.368)\\
 PD1 Test Orbit & (35.5, 7.80, 37.8) & (-0.0385, 0.272, 0.231)\\
\hline
\end{tabular}
\end{minipage}
\end{table}

For PD1, PD2 and PD3, and each of 15 values of the baseline distance
$s_{0\rm b}$, 280 optimisation attempts were made, each involving
Metropolis annealing for 24,000 simplex deformations, from an initial
temperature of 0.5$\,$dex. The starting distances were constrained by
the penalty function $p_s$ (eq.~\ref{eq:defsps}) with $\beta = 10^6$
and $\delta s = 0.5\kpc$, so the Metropolis algorithm could only
explore a narrow band in $s_0$. After 40 optimisation attempts from a
given starting point $\chi_{\rm guess}$, the starting point was
updated to the end point of the most successful of these
optimisations, and annealing recommenced from a high temperature. In
total 6 of these updates were performed. With these parameters, a
search at a single distance completes in 12 CPU-hours on 3GHz
Xeon-class Intel hardware. PD4, PD5 and PD6 follow almost the same
schema, except that 48,000 simplex deformations were formed for each
of 60 attempts at the same $\chi$, which was updated 6 times. A single
distance in the latter case took 36 CPU-hours to search: the
computational load scales linearly with the number of deformations
considered.

\begin{table}
 \centering
 \begin{minipage}{85mm}
  \caption{Configurations for tests of the method.}
  \label{testtab}
  \begin{tabular}{@{}lccccc@{}}
  \hline
  Set& $\delta v_{\parallel,{\rm max}}/\!\kms$ & $\delta v_{\parallel,{\rm min}}/\!\kms$ &
  offset $v_\parallel/\!\kms$ \\
 \hline
  PD1 & $2\times 10^{-3}$ & $2\times 10^{-3}$ & 0 \\
  PD2 & 4 & 0 & 0\\
  PD3 & 6 & 2 & 0\\
  PD4 & 10 & 10 & 0\\
  PD5 & 6 & 2 & 2\\
  PD6 & 15 & 10.5 & 10\\
  PD7 & 4 & 0 & 0\\
\hline
\end{tabular}
\end{minipage}
\end{table}

\figref{pd1} shows the results obtained with PD1, which has very small
error bars. The diagnostic function $D'$ has a smooth minimum that
pinpoints the distance $s_0=47.4\kpc$ to the starting point with an uncertainty of
$\sim 0.2\kpc$. The
Metropolis optimisation can significantly reduce $D'$ by tweaking the
input track only when $s_0$ is close to the truth. The depth of the
minimum indicates the numerical noise floor for this particular
problem, and no significant scatter is seen between successive runs.
 We do expect the noise floor to be slightly different for
PD2--PD6 because both the underlying orbits and the input are somewhat
different. \figref{pd1-vs-truth} shows that the reconstructed
solution at the minimum overlies the input orbit almost exactly. We
conclude that when the error bars are as small as in PD1, only one
orbit is consistent with the data.

\begin{figure}
\centerline{\psfig{file=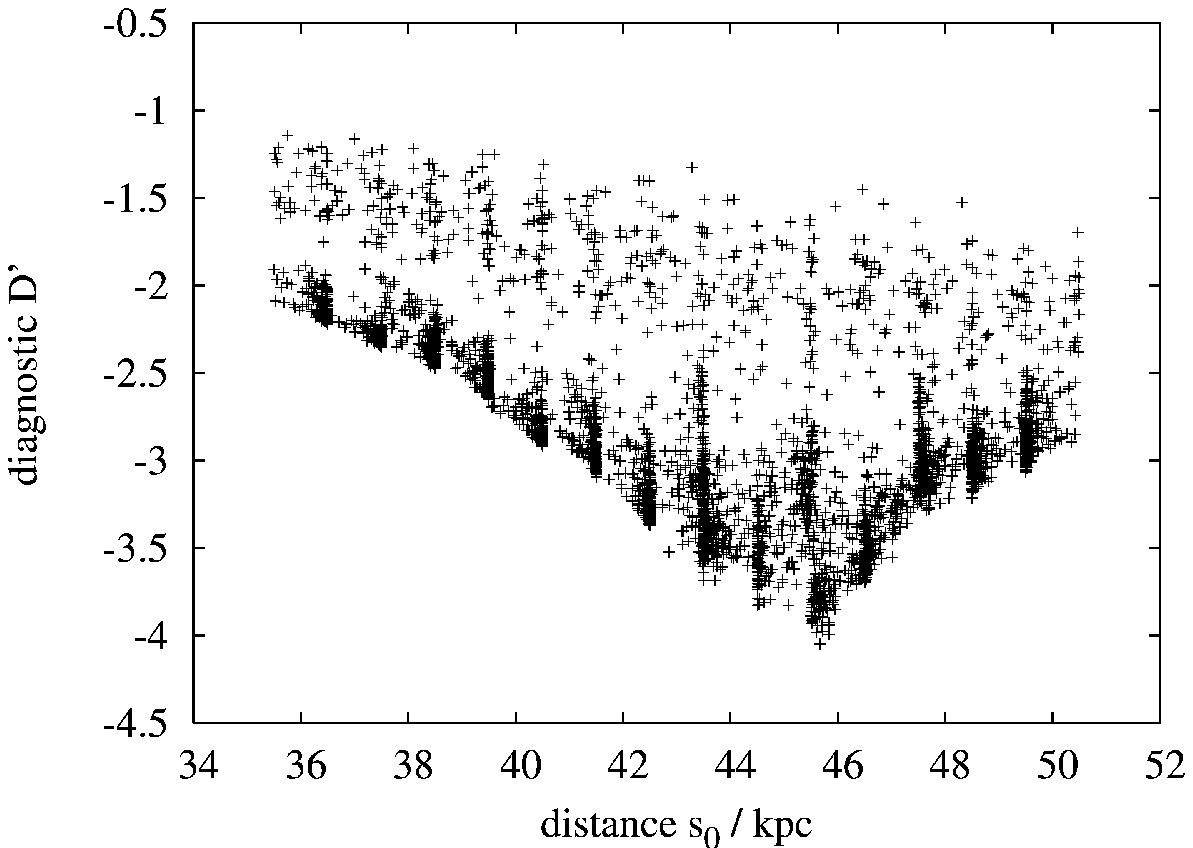,width=.93\hsize}}
\centerline{\psfig{file=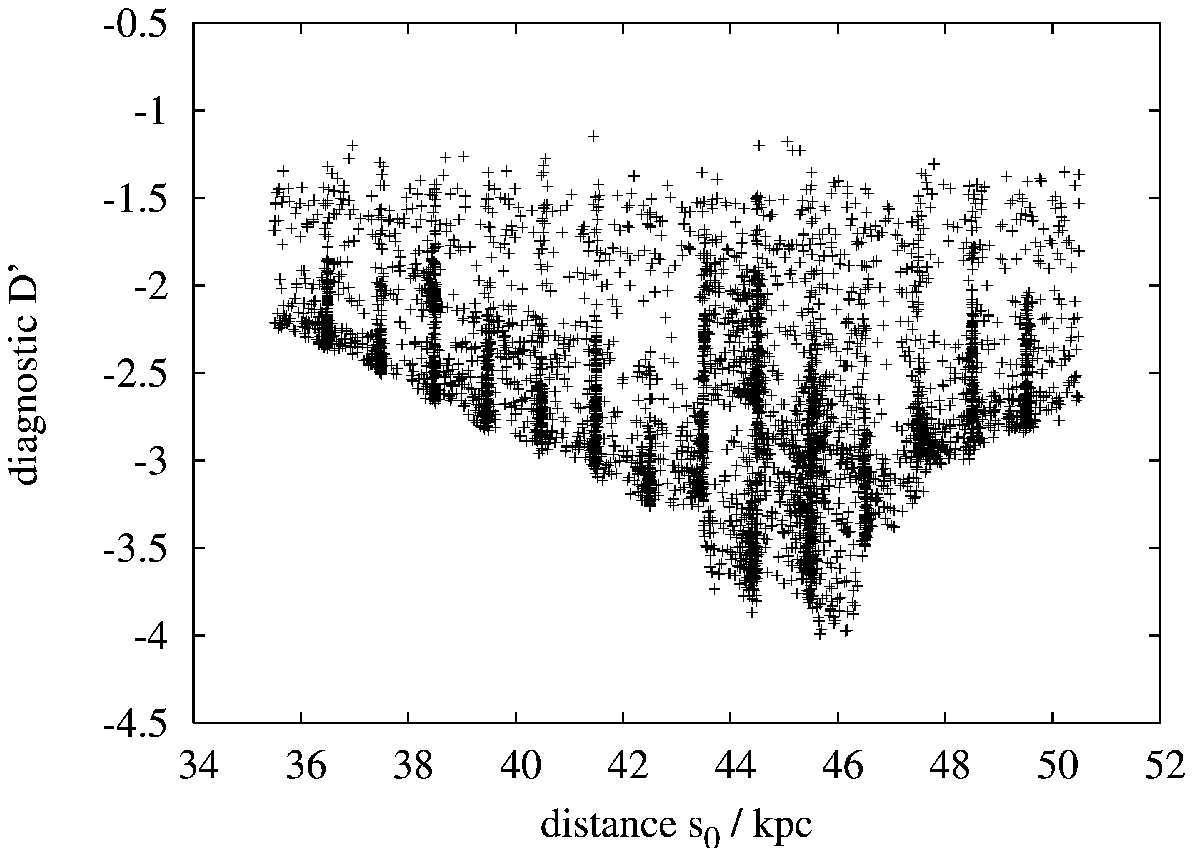,width=.93\hsize}}
\centerline{\psfig{file=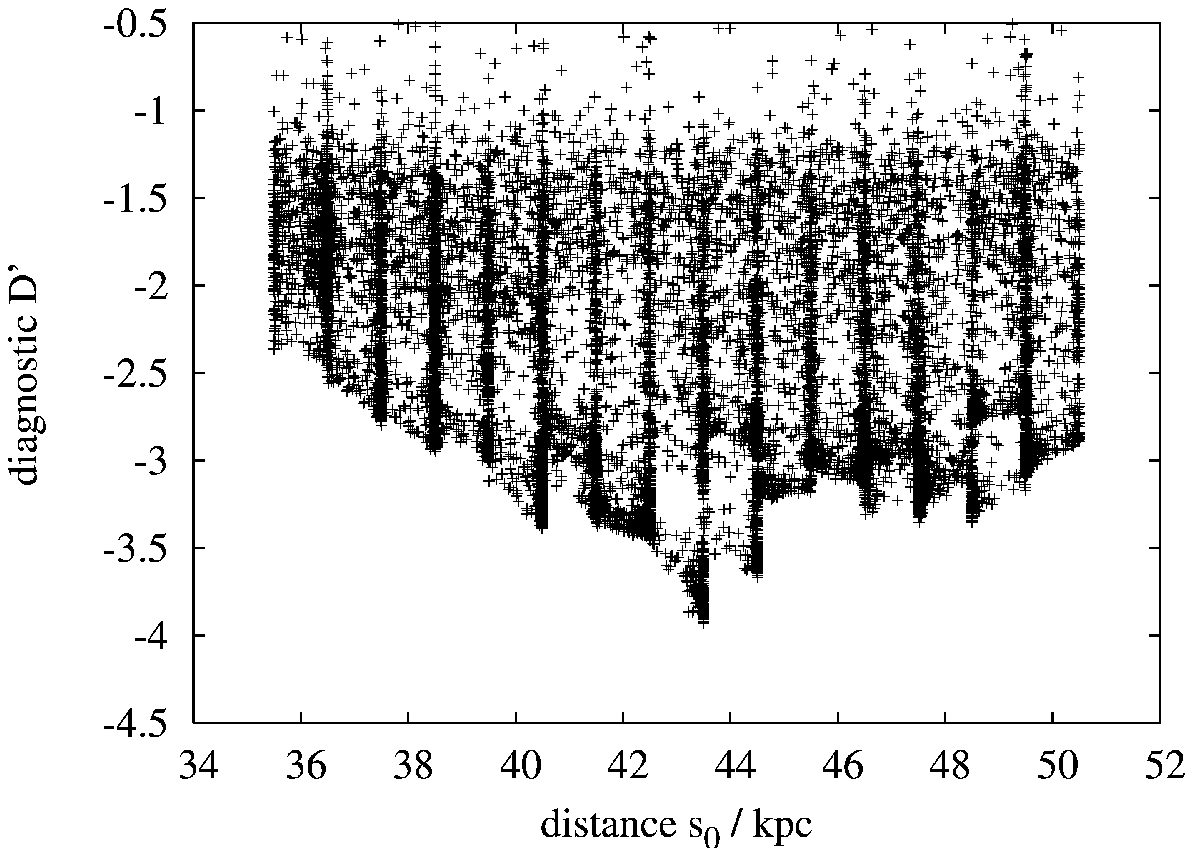,width=.93\hsize}}
\caption{The same as \figref{pd1} but for input data sets PD2 (upper),
  PD3 (middle) and PD4 (lower). In PD2 and PD3, the noise floor $D'
  \sim -4$ is reached only for starting distances in the range
  $44-46\kpc$. In PD4 we find an isolated good solution at $43\kpc$, but
  in contrast to what happens with datasets  PD2 and PD3, 
  as we change distance the value of $D'$  oscillates 
  around $\sim -3.25$ for most of the range, rather than varying smoothly.
  This behaviour
  arises because  the volume of parameter space that has to be searched is
  large on account of the largeness of the velocity errors. Such an extensive
  volume of parameter space cannot be exhaustively searched with the allocated
  computational resource. Only orbits with
  $s_0 < 40\kpc$ could be excluded with confidence.}
 \label{pd234}
\end{figure}

\begin{figure}
\centerline{\psfig{file=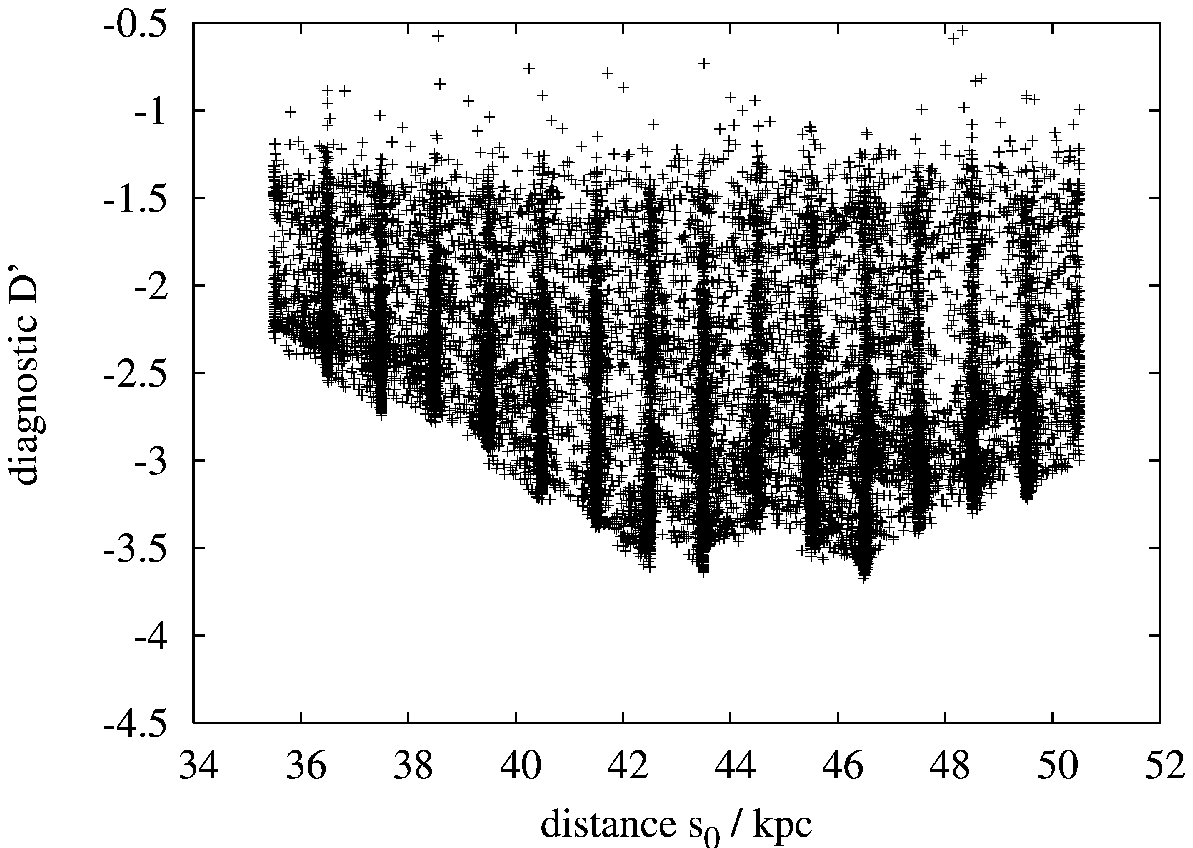,width=.93\hsize}}
\centerline{\psfig{file=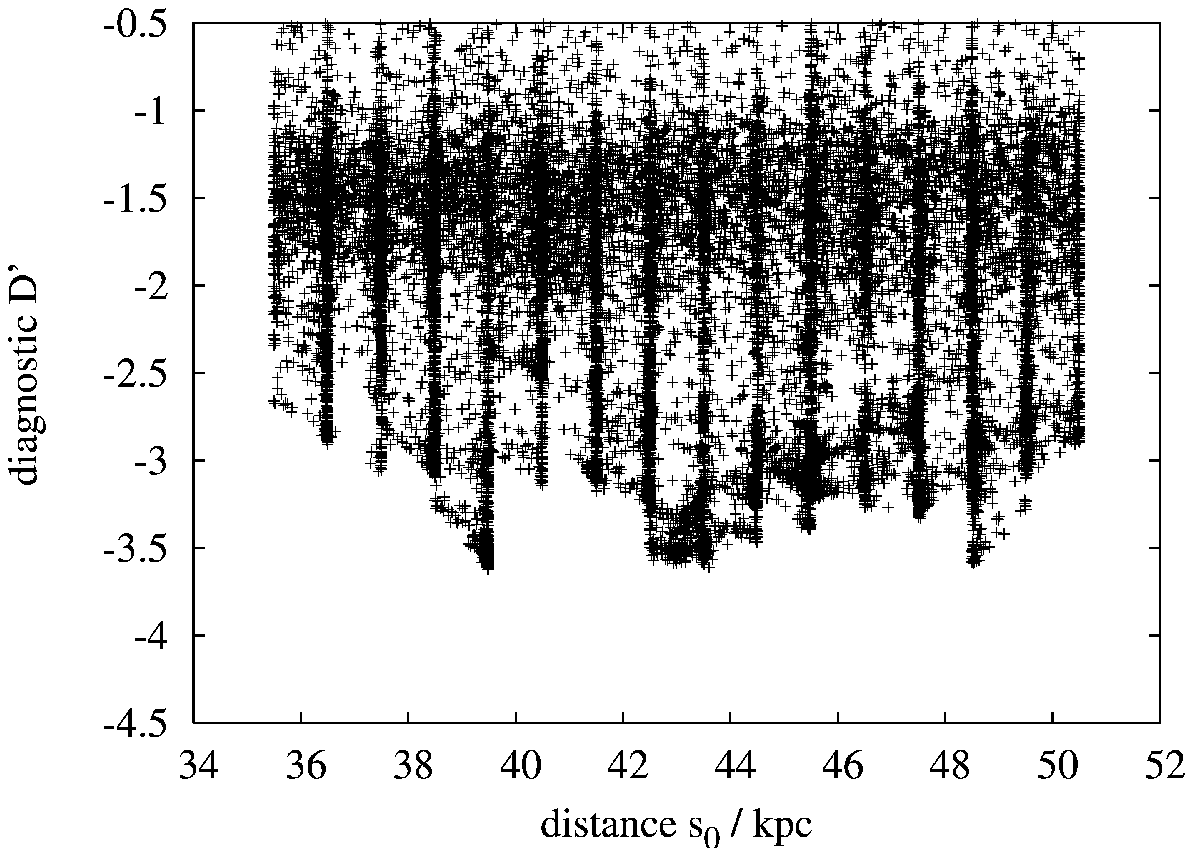,width=.93\hsize}}
\caption{The same as \figref{pd234} but for input data sets PD5 (upper),
PD6 (lower). The noise floor $D \sim -3.5$ is now approached over a wider range
in $s_0$: $42-47\kpc$ in PD5 with some confidence, and $38-48\kpc$ in PD6 with little confidence.
As the  errors increase, the search becomes a more arduous task, and patchy
performance of this task is reflected in the rough
bottoms to the graphs.}
\label{pd56}
\end{figure}

\begin{figure}
\centerline{\psfig{file=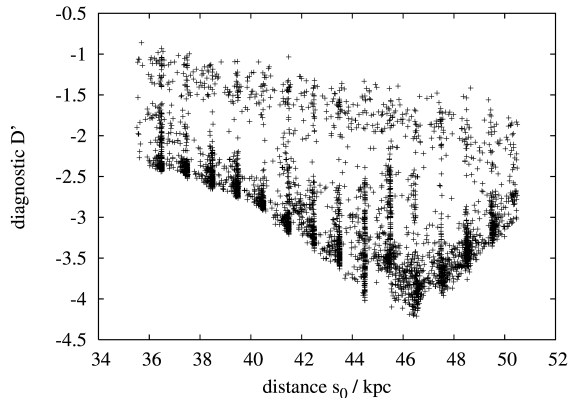,width=.93\hsize}}
\caption{The same as \figref{pd56} but for input data set PD7. The results are very similar
to those of PD2 and PD3 (\figref{pd234}), with a noise floor $D' \sim -4$ and solutions acceptable only in the
range $44-46\kpc$.}
\label{pd7}
\end{figure}

\begin{figure}
\centerline{\psfig{file=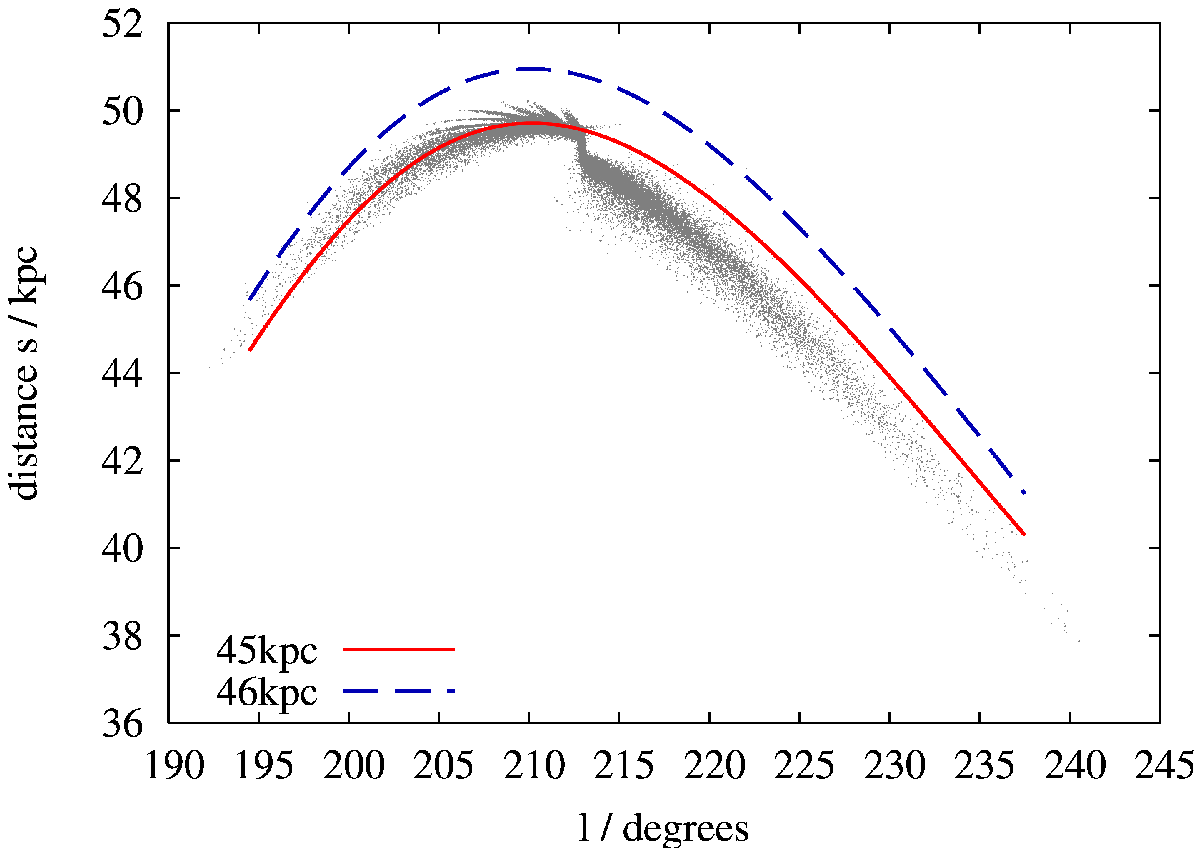,width=.93\hsize}}
\centerline{\psfig{file=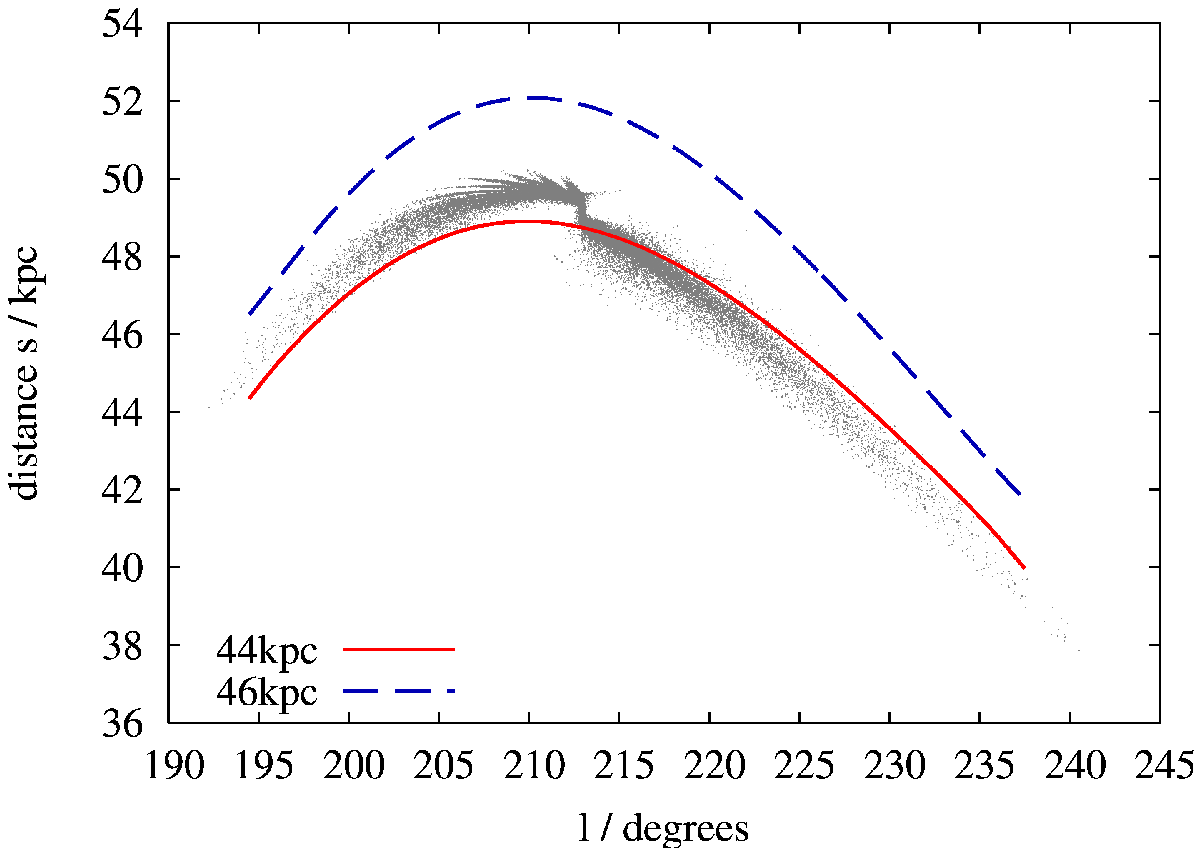,width=.93\hsize}}
\centerline{\psfig{file=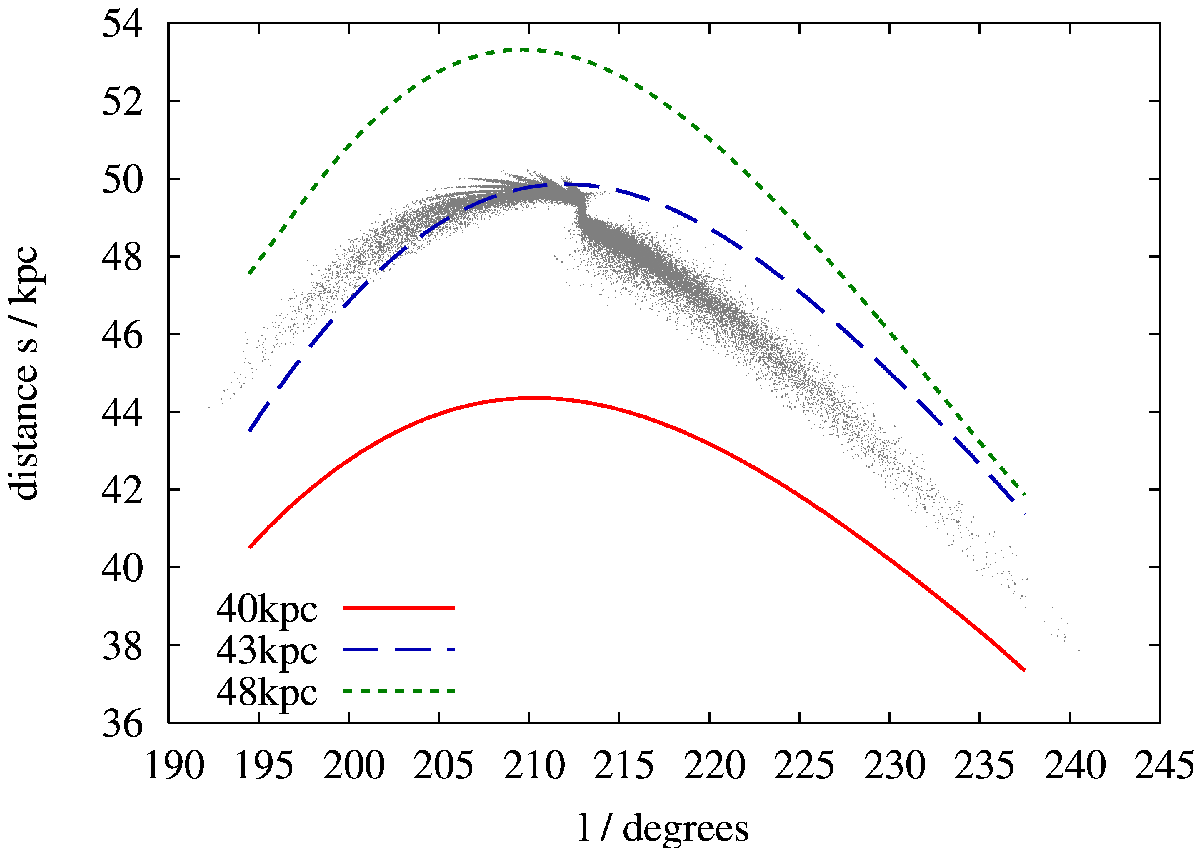,width=.93\hsize}}
\caption{Heliocentric distance for selected reconstructed orbits
from PD2, PD3 and PD4. The tracks selected are those with lowest $D'$ at
distances for which $D'$ approaches the noise floor in \figref{pd234}.}
\label{pd234-dist}
\end{figure}

\begin{figure}
\centerline{\psfig{file=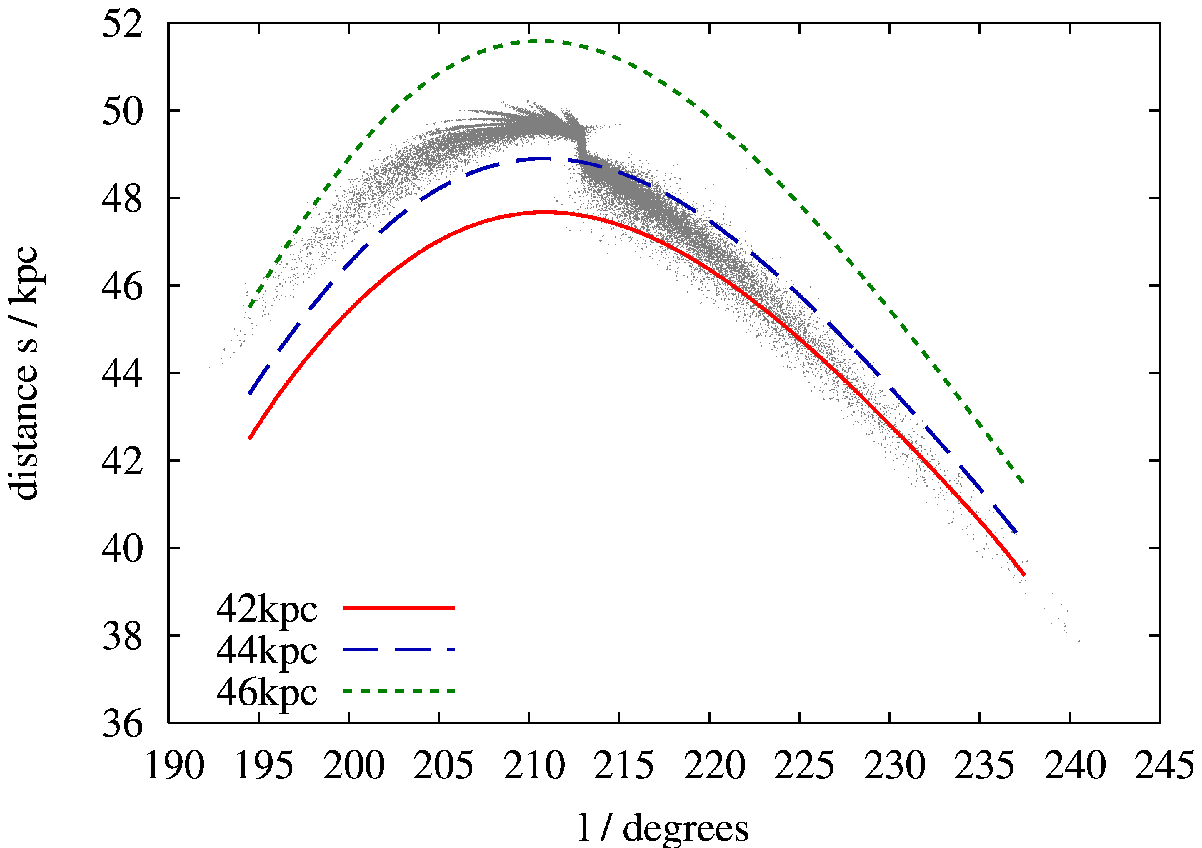,width=.93\hsize}}
\centerline{\psfig{file=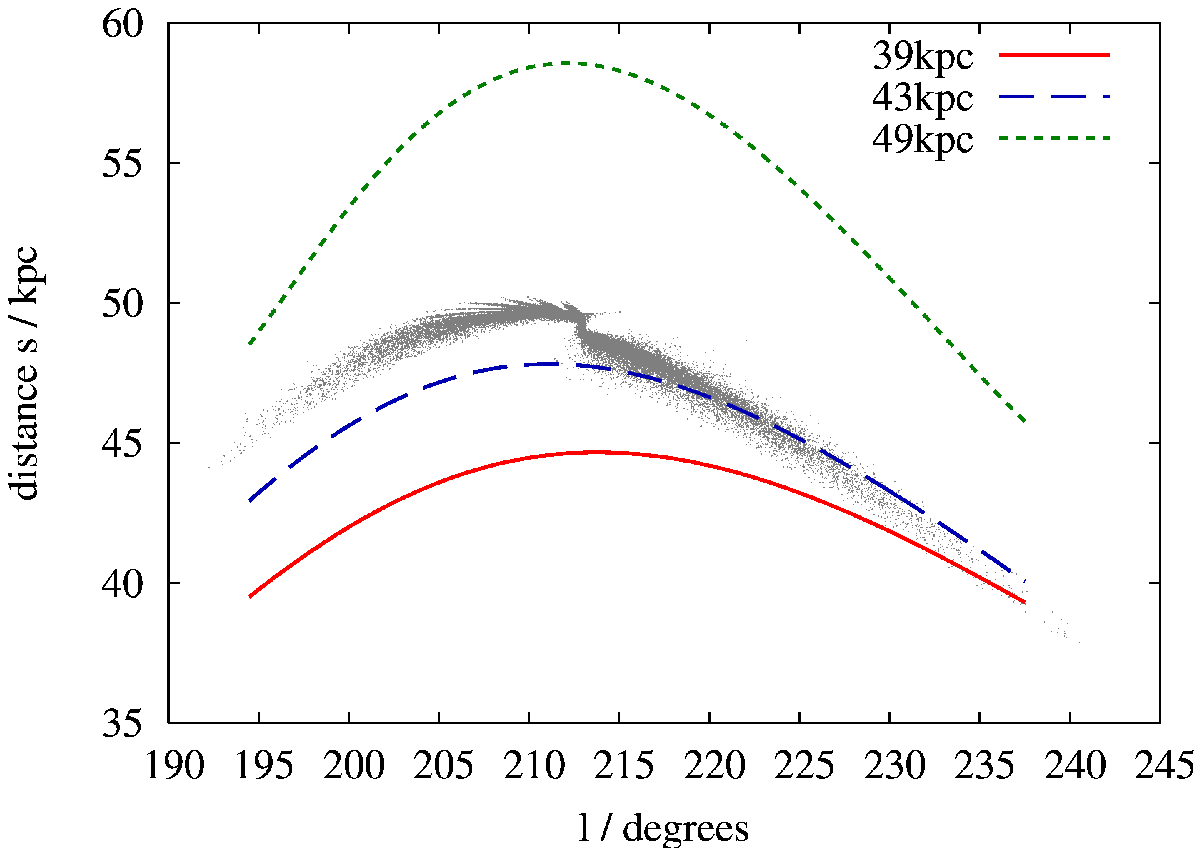,width=.93\hsize}}
\centerline{\psfig{file=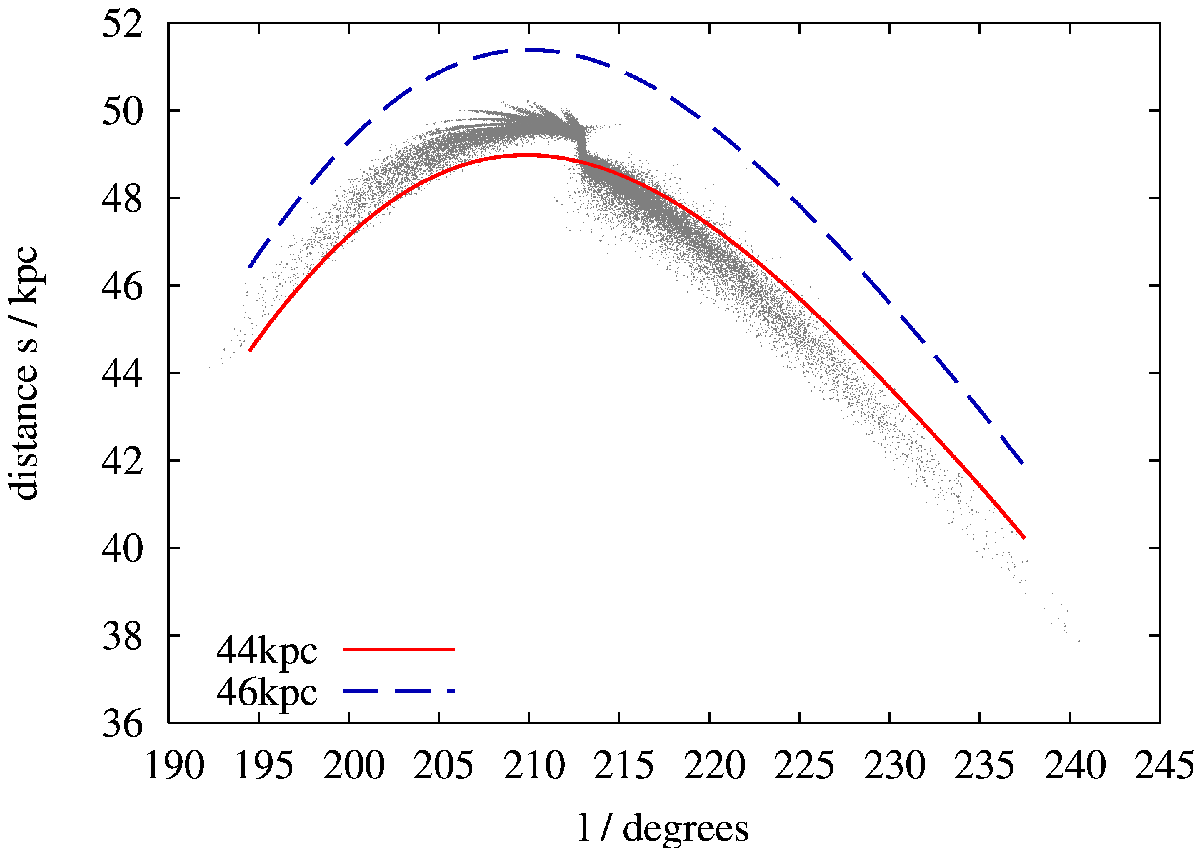,width=.93\hsize}}
\caption{As \figref{pd234-dist} except for data sets PD5, PD6 and PD7.}
\label{pd56-dist}
\end{figure}

\begin{figure}
\centerline{\psfig{file=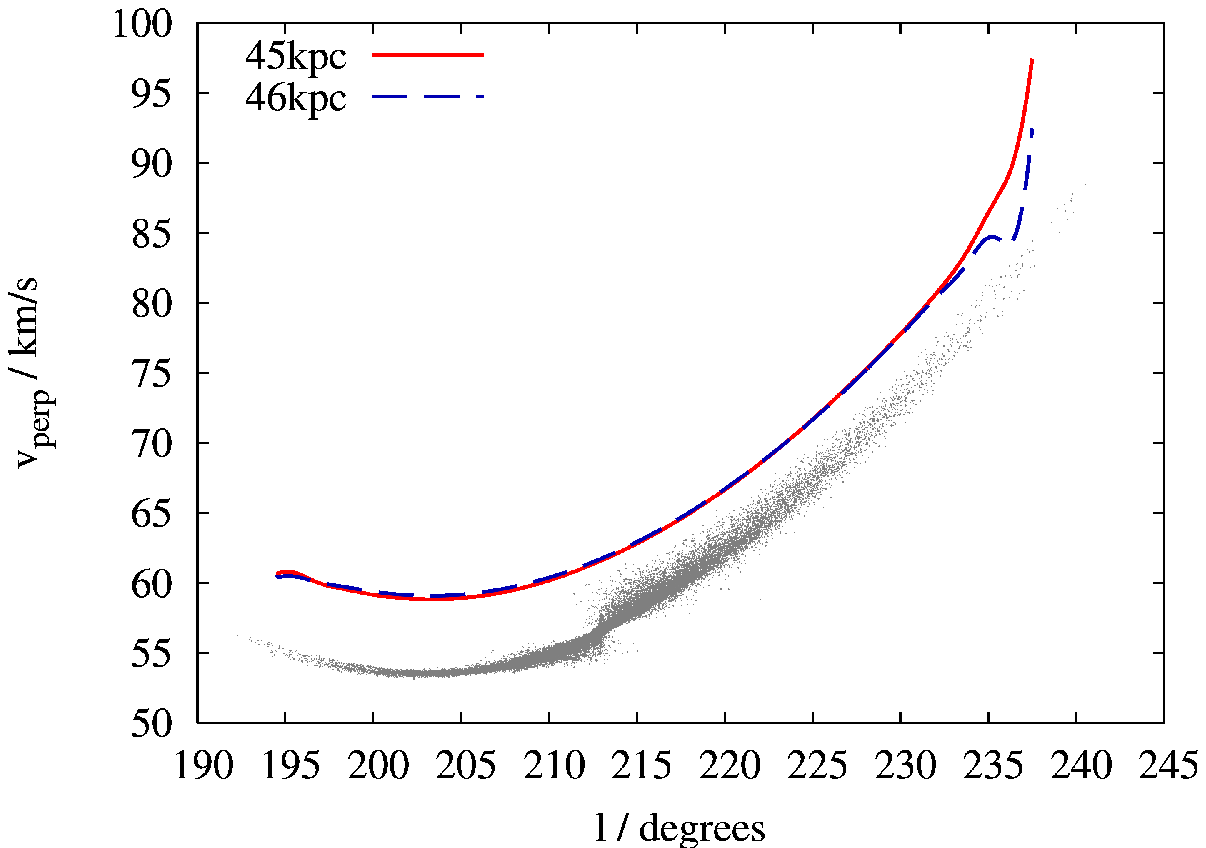,width=.93\hsize}}
\centerline{\psfig{file=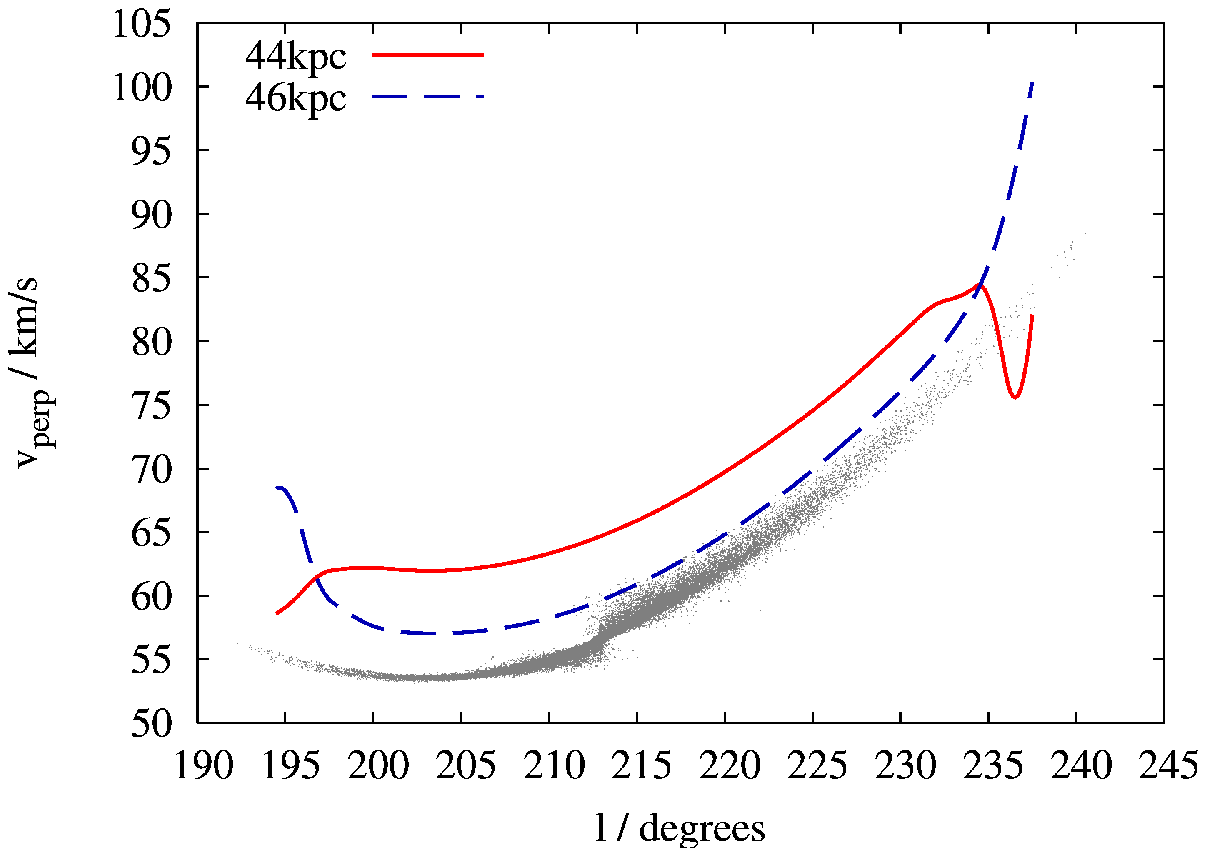,width=.93\hsize}}
\centerline{\psfig{file=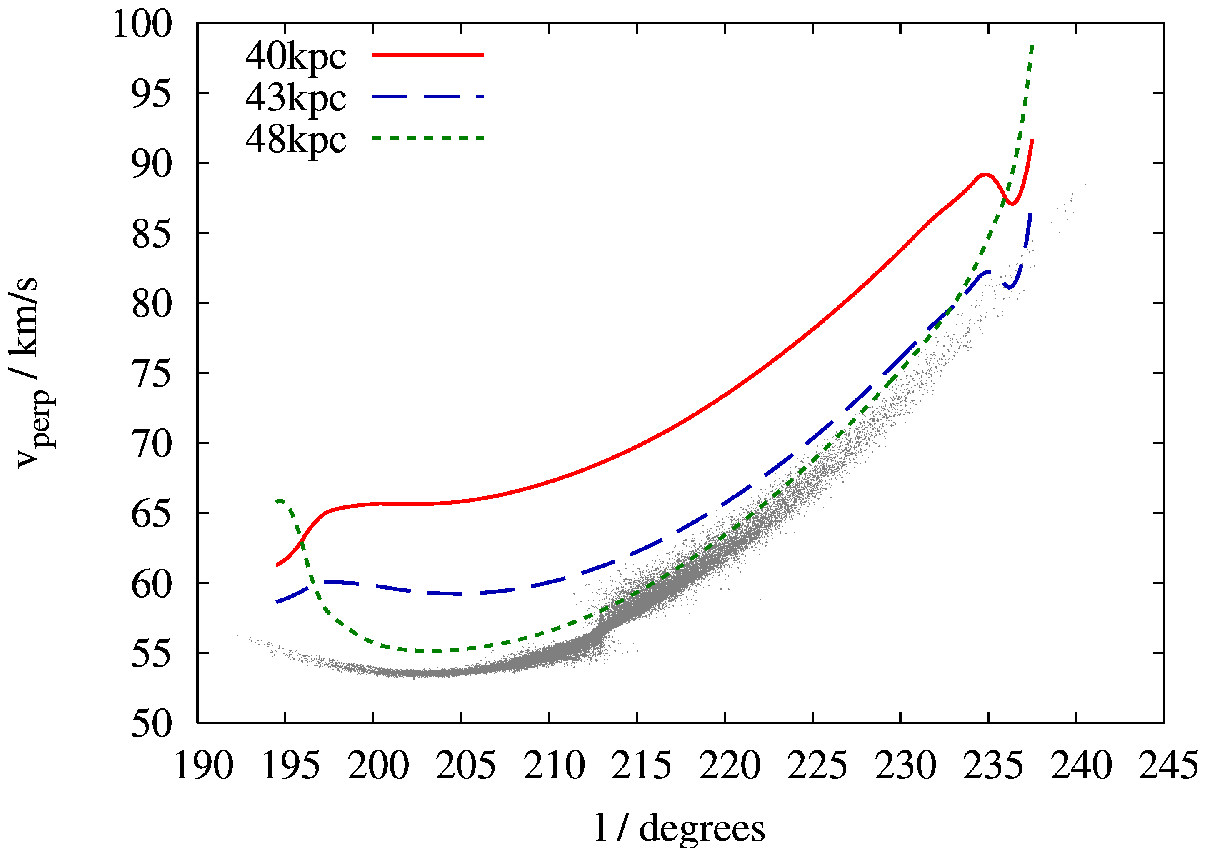,width=.93\hsize}}
\caption{Tangential velocities for selected reconstructed orbits from
PD2, PD3 and PD4. The tracks selected are the same as in \figref{pd234-dist}.}
\label{pd234-vt}
\end{figure}

\begin{figure}
\centerline{\psfig{file=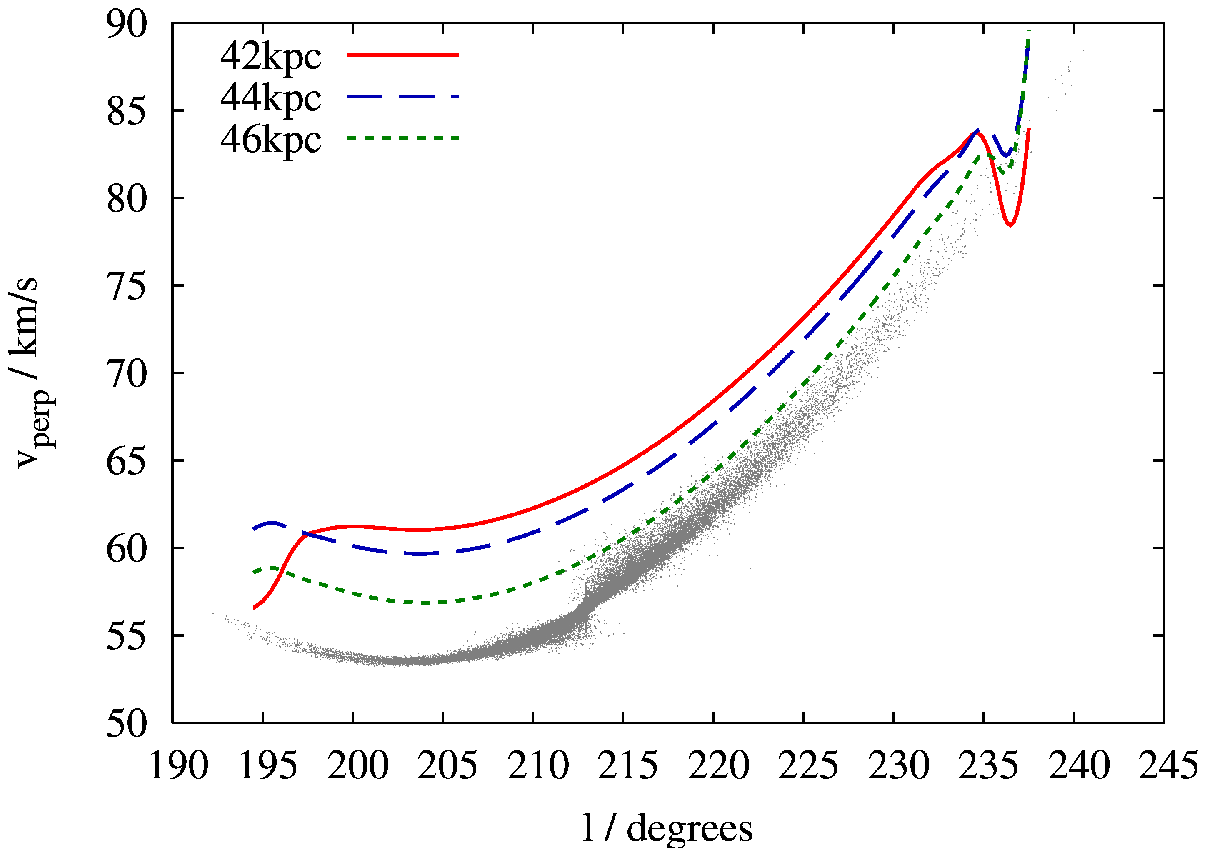,width=.93\hsize}}
\centerline{\psfig{file=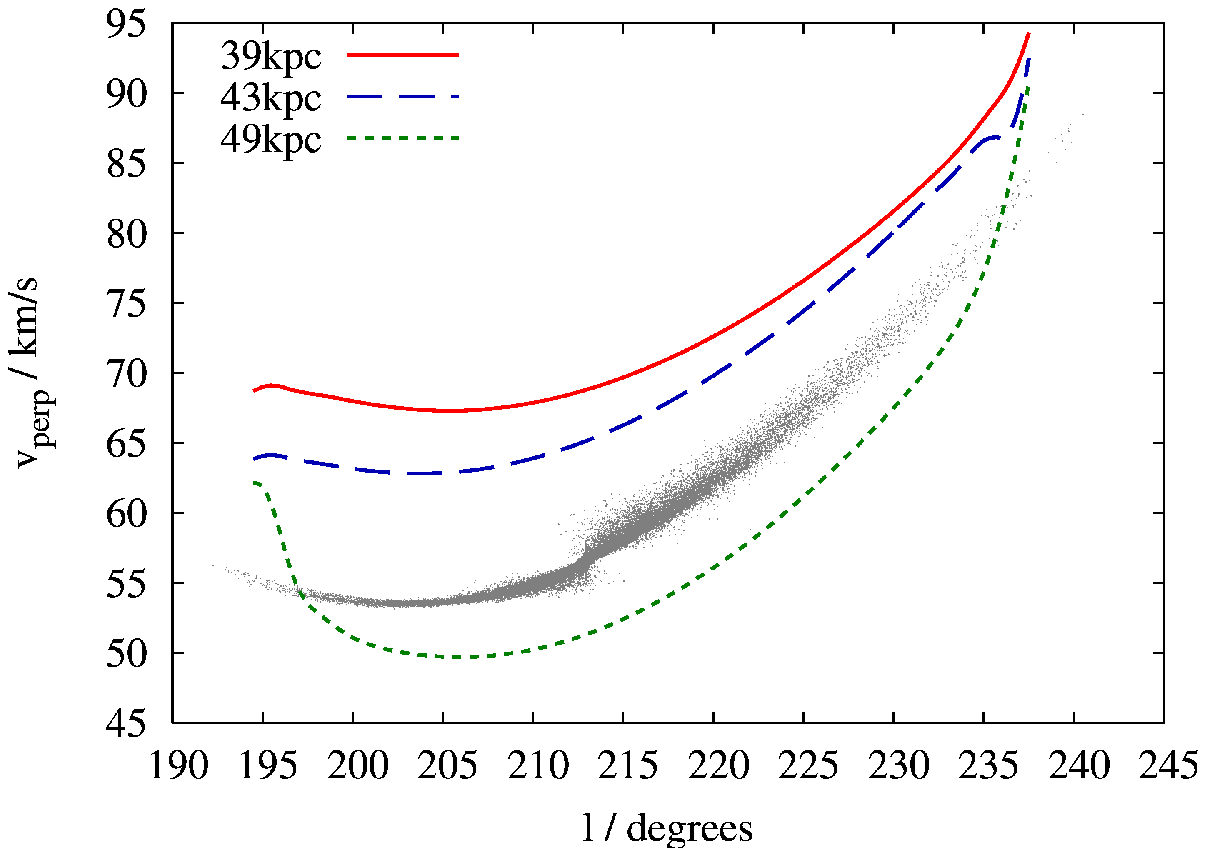,width=.93\hsize}}
\centerline{\psfig{file=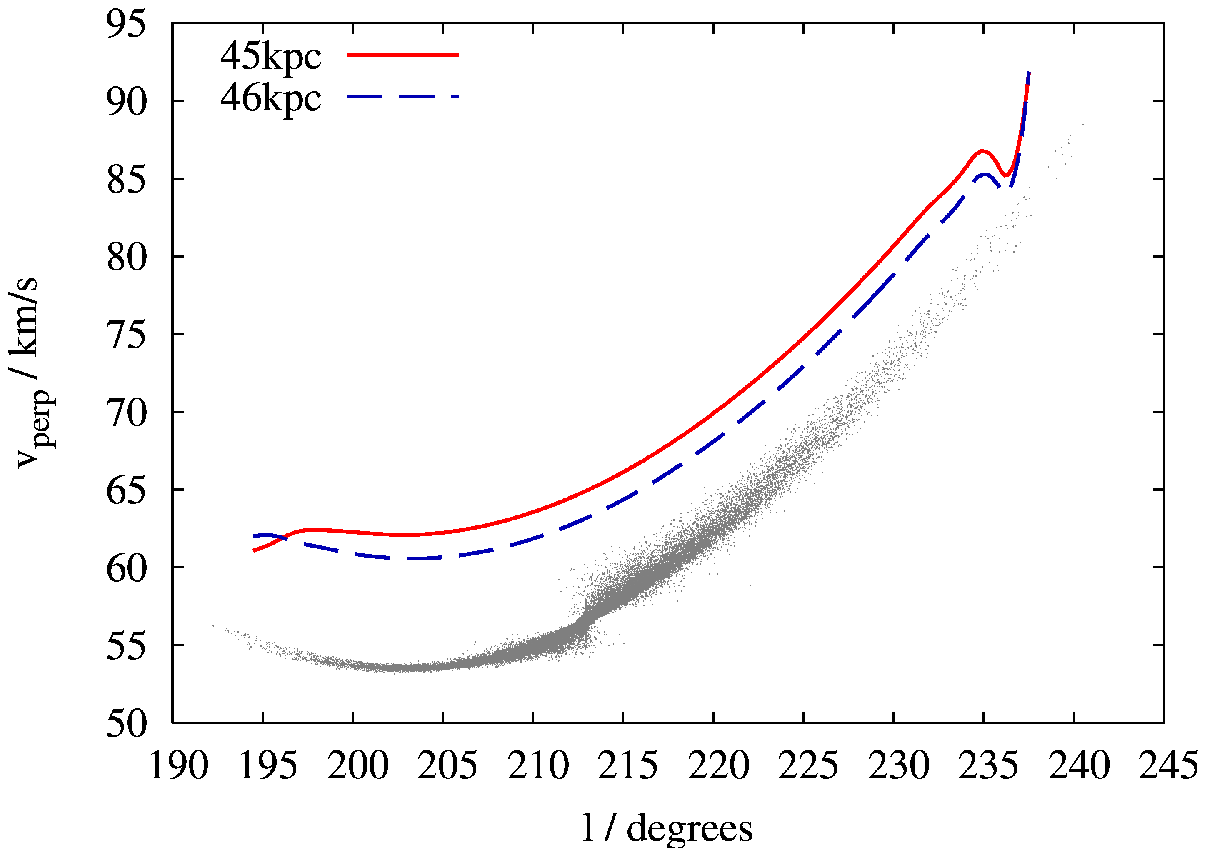,width=.93\hsize}}
\caption{As \figref{pd234-vt}, except for data sets PD5, PD6 and PD7.
The tracks selected are the same as in \figref{pd56-dist}.}
\label{pd56-vt}
\end{figure}

\figref{pd234} shows the results obtained with PD2, PD3 and PD4. For
PD2 and PD3, which have small to moderate error bars, the Metropolis
algorithm reduces $D'$ to the noise floor only for $s_0$ in the range
$(44,46)\kpc$. The scatter between runs is small, $\sigma_{D'} \sim 0.1$.
 \figref{pd234-dist} and \figref{pd234-vt} show
the reconstructed distances and tangential velocities associated with
the best two solutions found at $44$ and $46\kpc$. With PD2, these reconstructions
provide a distance estimate to the stream that is, at worst, $2\kpc$ in
error, and a $\vperp$ estimate that is at worst $5\kms$ in error. With PD3, the
reconstruction is, at worst, $3\kpc$ and $10\kms$ in error. For many sections of
the orbits, the errors are less than stated. Thus the method can
identify orbits consistent with the stream, and reject those that are
inconsistent with it. 

For PD4, which has large error bars comparable to those for velocity data
from the SDSS, scatter between repeat runs is much larger,
$\sigma_{D'} \sim 0.5$. Only distances $s_0 < 40\kpc$ can be
confidently excluded, although a high-quality solution near $s_0 \sim
43\kpc$ provides a reconstruction in error by at most $2\kpc$ in
distance and $5\kms$ in $\vperp$. \figref{pd234-dist} and
\figref{pd234-vt} shows us that ignoring this high-quality solution
would give reconstructed quantities in error by up to $5\kpc$ in
distance and $12\kms$ in $\vperp$. Increasing the errors associated
with the input clearly permits less satisfactory solutions to be
returned, and also decreases the ability of the search procedure to
find true orbits at particular distances, should they exist, by
increasing the volume of parameter space it has to search. The former
complaint is a physical statement about the limitations of the input data. The
latter complaint is an algorithmic one, which may be remedied by
providing the search with more computational cycles, or improving its
efficiency.

\figref{pd56} shows the results obtained from input sets PD5 and PD6
in which offsets were applied to the input velocities.  The
reconstructed distances and tangential velocities for interesting
tracks are shown in \figref{pd56-dist} and \figref{pd56-vt}. For PD5,
the scatter between runs is $\sigma_{D'} \sim 0.5$, and the distance
to the stream $s_0$ can be said to lie in the range $42-47\kpc$
with some confidence.  \figref{pd56-dist} shows that the stream
does indeed lie in this range, so the algorithm has successfully
corrected the small offset. The error in reconstructed distance and
$\vperp$ would be $3\kpc$ and $8\kms$ at worst.

PD6 demonstrates the limits of the method. On account of the large scatter in $D'$,
$\sigma_{D'} \sim 1$, we cannot identify the correct distance from
\figref{pd56}. We expect the distance range for permitted orbits to be wide
with this input, because the velocity error bars are
large. \figref{pd56} illustrates another problem of using large error
bars: the search becomes harder because the parameter space to be
searched is much larger.  Consequently the plots of the results have
rough bottoms, where the algorithm has failed to reach consistent
minima for searches at adjacent distances.  This can be remedied by
re-running the search with more deformations and more iterations, and
may be addressed in future upgrades to the search procedure.

PD7 demonstrates that the accuracy of the method is not necessarily
significantly degraded when the number of velocity data points is
substantially lower than the number of positional data points, as might be
expected from a stream for which the only available radial velocities are
those of giant stars.  The results for PD7 are shown in \figref{pd7} and are
directly comparable to those of PD2 and PD3 (\figref{pd234}). In particular,
the range of allowed distances, $44-46\kpc$, is the same and the
reconstructed orbits show comparable accuracy (Figs.~\ref{pd56-dist} and
\ref{pd56-vt}).

\begin{figure}
\centerline{\psfig{file=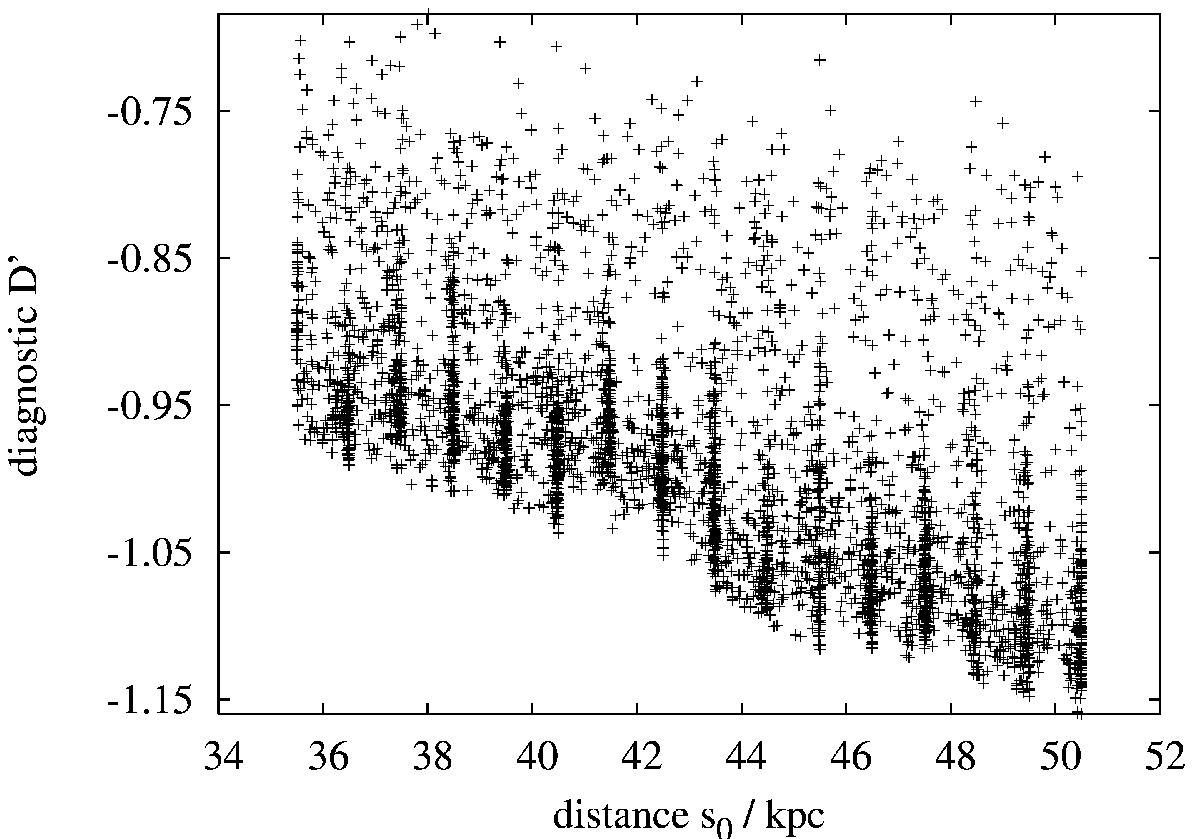,width=.93\hsize}}
\centerline{\psfig{file=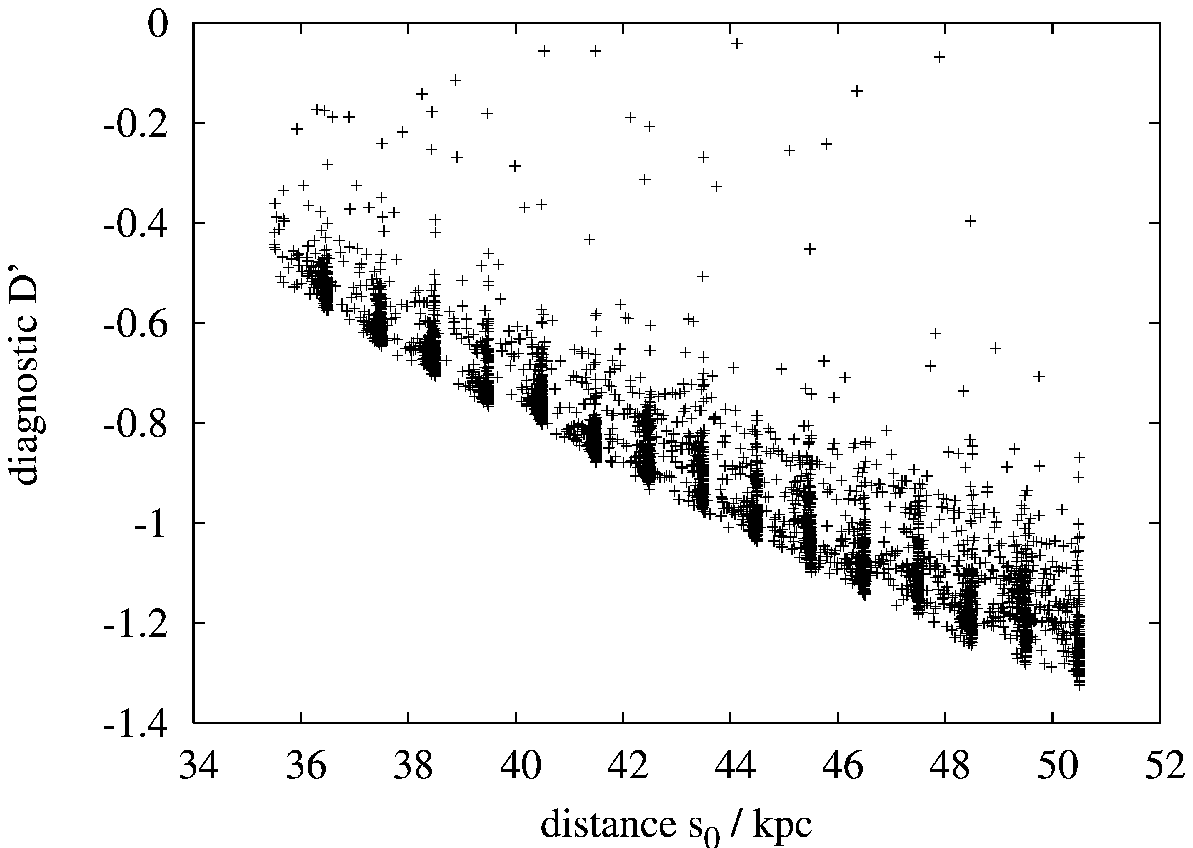,width=.93\hsize}}
\caption{As the top panel of \figref{pd234}, except the
  reconstruction (using PD2) takes place in a Keplerian potential (top panel), and
  (bottom panel) a potential with $\Phi(r) \propto r$. In both cases,
  the constants are set to generate approximately the same passage time
  along the stream as in Model II. Comparing with
  \figref{pd234} shows the values of $D'$ are very poor, demonstrating
  that no orbits can be found in these potentials, so the latter can be
  excluded.}
\label{kepler}
\end{figure}

\begin{figure}
\centerline{\psfig{file=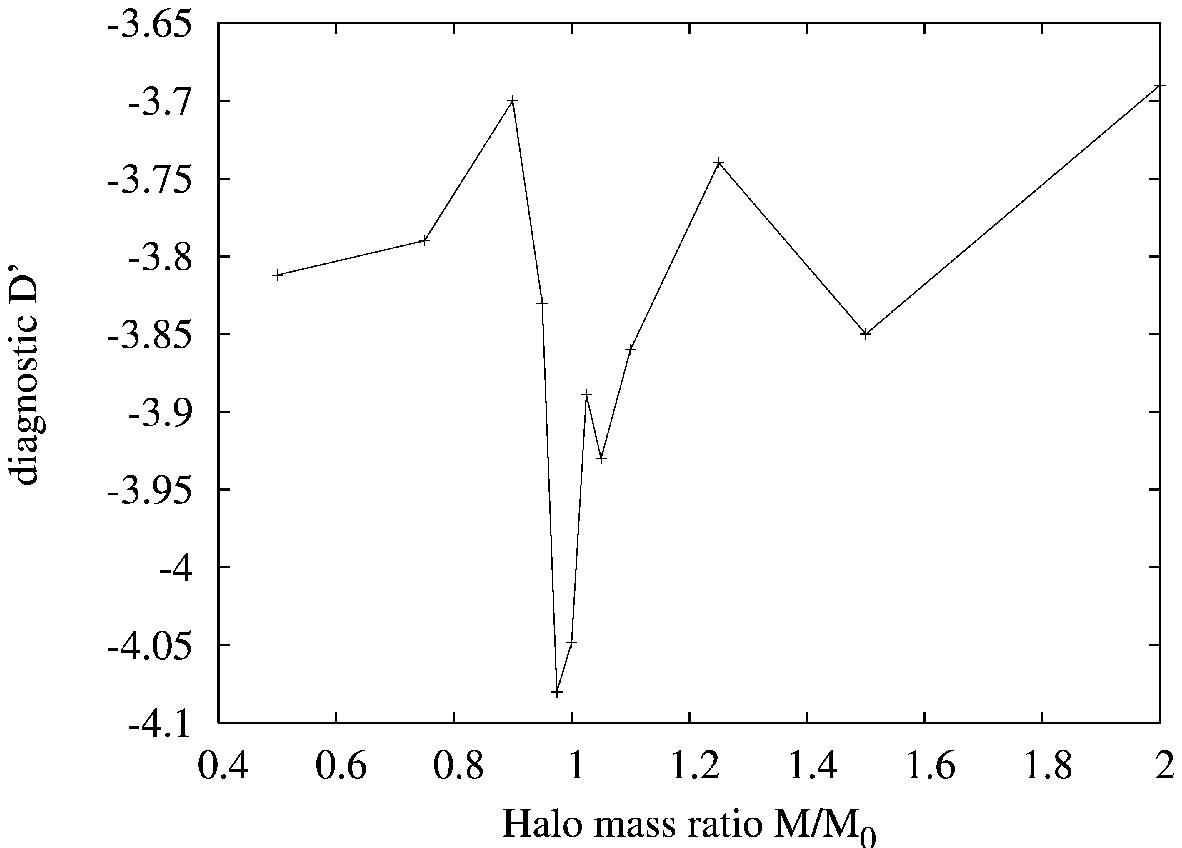,width=.93\hsize}}
\centerline{\psfig{file=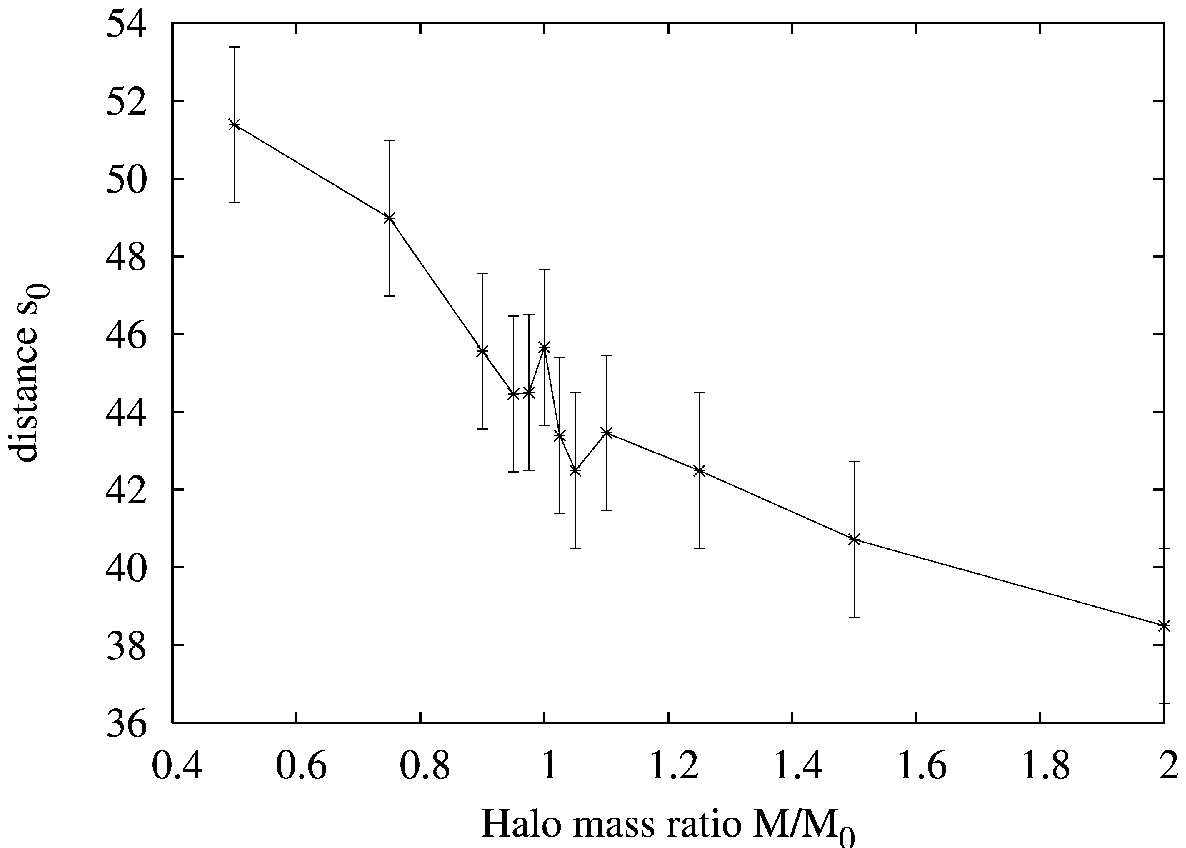,width=.93\hsize}}
\caption{The upper panel shows the minimum value of the diagnostic for
attempted reconstruction of the PD2 input, versus the halo mass ratio of the Dehnen-Binney
potential in which the reconstruction has been attempted. $M_0$ here is the value of the
halo mass in the Model II potential. The lower panel shows the characteristic distance,
$s_0$, of the best solution, versus halo mass ratio. The error bars represent the
approximate uncertainty in the recovered result.}
\label{changepot}
\end{figure}

\section{The effect of changing the potential}\label{sec:potential}

Paper I demonstrated the ability of orbit reconstruction to diagnose the
Galactic potential with astonishing precision when the track of an orbit on the sky
is precisely known. Here we investigate the ability of  orbit reconstruction
to identify the correct potential when the track has to
be inferred from realistic stream data. 

We use as our input the PD2 data set from Section~\ref{sec:test}.  The
top panel of \figref{kepler} shows the results of asking the algorithm
to find orbits in a Kepler potential with mass $M = 4.18 \times
10^{11} M_{\sun}$, which produces roughly the same passage time along
the stream as does Model II. The bottom panel shows the results obtained
using a potential of the form $\Phi(r) = r f_r$, which gives a rotation
curve of the form $v_c(r) = \sqrt{r f_r}$, with $f_r = 6.86 \times 10^2 (\kms)^2/\kpc$ again
chosen to produce the same passage time along the stream as does Model II.
These two potentials represent, respectively, relatively extreme falling
and rising rotation curve models. We do not offer them as realistic candidates
for the Galactic potential, but intend to demonstrate that model potentials with
approximately correct radial force, but incorrect shape, can be excluded using
this method.

The distance range considered in both cases spans
approximately $\pm 15$ percent of the true distance, which is the
uncertainty one might expect in distances obtained by photometry. Since all
values of $D'$ are several orders of magnitude larger than the
values one obtains with the correct potential, it is clear that no
orbits can be found at the distances considered, and both potentials
can be excluded.

The upper panel of \figref{changepot} shows the results of reconstructing an
orbit from the PD2 data set in a potential that differs from the Model II
potential used to define the tidal stream only in that the halo mass has been
varied by
the specified ratio.  This high-quality data set yields a sharp minimum in
$D'$ as a function of $s_0$ (\figref{pd234}) and the upper panel of
\figref{changepot} shows this minimum value of $D'$ as a function of assumed
halo mass. The minimum of $D'$ lies at $D'\lta-4$ when the mass used lies
within $\sim5$ percent of the true value and $D'>-3.85$ otherwise.

The lower panel of \figref{changepot} shows that the value of $s_0$ at
which $D'$ attains its minimum for given halo mass decreases
systematically as the halo mass increases. Insight into this behaviour
can be obtained by considering the discretised equation of
angular-momentum conservation
 \beq
 \Delta(sv_\perp)=-s{\d\Phi\over\d
  r_\perp}\Delta t =-s{\d\Phi\over\d r_\perp}{\Delta v_\parallel\over
  F_\parallel}, 
\eeq
 where $\Delta$ implies the change in a quantity
between successive data points. By expanding the left side to first
order in small quantities we can obtain an expression for $\Delta
v_\perp$. Summing the changes in $v_\perp$ along the track we have
 \beq\label{vpfirst}
v_\perp(\hbox{end})-v_\perp(\hbox{start})=-\sum\left({\d\Phi\over\d
    r_\perp}{\Delta v_\parallel\over F_\parallel}+v_\perp{\Delta
    s\over s}\right).
\eeq
 An independent equation for $v_\perp$ is
\beq\label{vpsecond}
 v_\perp=s{\Delta u\over\Delta t}=sF_\parallel
{\Delta u\over \Delta v_\parallel},
\eeq
 where we have used the radial equation of motion. Equations (\ref{vpfirst})
and (\ref{vpsecond}) yield independent estimates of
$v_\perp(\hbox{end})-v_\perp(\hbox{start})$. The right side of equation
(\ref{vpfirst}) yields an estimate that is independent of the scaling of
$\Phi$ but systematically decreases with increasing $s$, while the right side
of equation (\ref{vpsecond}) yields an estimate that is proportional to the
scaling of $\Phi$, but is almost independent of $s$, since $F_\parallel \propto
1/s$. When the machine is asked to reconstruct the orbit with $F_\parallel$
taken too small, it can change the right side of equation (\ref{vpfirst}) to
match the new value of the right side of equation (\ref{vpsecond}) by
increasing $s$. In this way, the discrepancies between the left sides of
equations (\ref{vpfirst}) and (\ref{vpsecond}), which contribute
substantially to the diagnostic $D'$, can be largely eliminated by increasing
$s$ as $\Phi$ is scaled down.

In principle this variation in reconstructed distance with the scaling of the
potential could be combined with photometric distances to constrain the
potential. Unfortunately, \figref{changepot} shows that
in the particular geometry under consideration,
even a $10$ percent distance error would produce a $\sim50$ percent error in
the estimate of the halo mass. Further work is required to discover what
effect the geometry of a particular stream has on its sensitivity to the
potential. Also of interest is whether simultaneously using multiple streams,
or streams with multiple wraps (such as the Sagittarius Dwarf stream), can
provide tight constraints on the potential.

\section{Conclusions}\label{sec:conclusions}

Paper I demonstrated the tremendous diagnostic power that is available if one
knows the track of an orbit on the sky and the associated line-of-sight
velocities. Tidal streams are made up of objects that are on closely related
orbits. In particular, they roughly delineate the underlying orbit, but they
do not do so exactly. We have presented a technique for identifying an
underlying orbit and thus predicting the dynamical quantities that have not
been observed, such as the distances to the stream and the proper motions of
its particles.

The technique involves defining a space of tracks on the sky and sequences of
line-of-sight velocities that are consistent with the observational data,
given the observational errors and the extent to which streams deviate from
orbits.  The equations of Paper I are used to determine a candidate orbit for
each track, and then the extent to which the candidate satisfies the
equations of motion is quantified -- in Paper I only violations of energy
conservation along a candidate were quantified.  This diagnostic quantity is
then used to search for tracks that could be projections of orbits
in the Galactic potential. In practice the search is conducted for several
possible distances to a fiducial point on the stream. If constraints on this
distance are available from photometry, the computational effort of the
search can be reduced by narrowing the range of distances for which searches
need to be conducted.

We have taken the Galaxy's gravitational potential and the solar
velocity with respect to the Galactic centre to be known. Our tests revolve
around an N-body model of the Orphan Stream \citep{orphan}. We
show that for this stream, which is $\sim40^\circ$ long and $0.3^\circ$
wide at its ends, distances to and tangential velocities of points
on the stream can be recovered to within $\sim 2 \kpc$ and $\sim5
\kms$, respectively, if radial velocities accurate to $\sim 1\kms$ are
measured. As the errors in the measured radial velocities increase,
the space of tracks that must be considered grows bigger and the
search for acceptable orbits becomes more laborious. Moreover, the
range of distances for which acceptable orbits can be found
broadens. However, even with errors in radial velocities as large as
$\pm 10\kms$, the uncertainties in the recovered distances are no
greater than $\sim 10$ percent and the recovered tangential velocities are
accurate to better than $\sim 20$ percent. Zero-point errors in the input
velocities that are reflected in appropriately wide error bars broaden
the range of acceptable orbits but do not skew the results.

We have shown that the method maintains its accuracy even when very few
radial velocity points are used to define the input, as might be necessary
when radial velocities can only be measured for giant stars. In our tests,
comparable results were obtained from pseudodata based on only three velocity
measurements and from pseudodata based on fifteen velocity measurements.
Indeed, the results obtained with three accurate velocity measurements were
significantly superior to those obtained with fifteen lower-quality
measurements. Naturally, exactly how many points are required to provide
well-determined input will depend on the shape of the radial velocity curve
along the stream in question.

We expect the accuracy of reconstructions to depend on the geometry
of the problem in hand. In particular, we expect streams at apocentre, 
where families of orbits are compressed both on the sky and
in radial velocity, to yield poorer results than streams away from apocentre.
Unfortunately, streams are most likely to be discovered at apocentre
because both orbital compression and low proper motions around
apocentre lead to a high density of stars at apocentre. We expect streams
that are relatively narrow to produce more accurate results, because the permitted
deviation of the orbit from the stream is then low. We also expect
to have more difficulty reconstructing orbits from streams that contain
a visible progenitor, since the potential of the progenitor will cause orbits
in the progenitor's vicinity to differ materially from orbits in the Galaxy's
underlying potential.

Paper I suggested that it should be possible, if sufficiently accurate
input is provided, to constrain the Galactic potential, since the
wrong potential will not admit an acceptable orbit. We have tested
this possibility for input with realistic errors. We find that two
potentials of significantly different shape, the Kepler potential and
$\Phi(r) \propto r$, are clearly excluded. We have also tested for
changes in scaling of an otherwise correctly-shaped potential, by
varying the mass of the assumed dark halo around the value used to
make the pseudo-data. In this case, we find the correct potential is
identified, with the diagnostic quantity generally worsening as the
halo mass moves away from its correct value by more than $\sim5$
percent. We further find a consistent relationship between the
reported stream distance and the halo mass with which the
reconstruction takes place. Although the reported distance is only
weakly dependent upon halo mass, this does open the possibility of
using alternative distance measurements, such as photometric
distances, in conjunction with these techniques to constrain the
Galactic potential. Further work is necessary to determine a full
scheme to recover parameters of the potential from stream data.  Also
in question is the extent to which simultaneous reconstruction of
multiple streams, and reconstruction of streams with multiple wraps
around the Galaxy, might provide stronger constraints on the Galactic
potential than the short section of a single wrap that we have
considered.  It may also prove possible to refine the other main
assumption of our scheme, the location and velocity of the Sun.

It is instructive to compare our method of finding orbits of progenitors with
the traditional N-body method. First our method explores each orbit at a tiny
fraction of the computational expense of N-body modelling, so it is feasible
to automate the search of orbit space. Moreover, the search lends itself to
parallelisation.  Whereas only a successful attempt to model a stream with N
bodies yields an interesting conclusion, our method can show that no orbit is
consistent with a given range of distances. 

There are several directions in which this work could be profitably extended.

\begin{itemize}

\item[1.]
There is scope for a
powerful synergy between our method and N-body modelling: our method is first used to
identify a likely orbit and this orbit then provides initial conditions
for an N-body simulation, which reveals the offset between the
progenitor's orbit and the stream. This knowledge would enable the bow-tie
region to be made narrower. Finally our method is used again to determine the
orbit with still higher precision.

\item[2.]  Currently VLBI observations of masers yield trigonometric
parallaxes for of
order two dozen sources located at several kpc from the Sun that are
accurate to several percent \citep{Reid09}, and Gaia will yield results of
similar precision for several million stars. The method discussed here
promises distances of slightly higher precision to sources that lie $50$ to
$100\kpc$ from the Sun. These ``geometrodynamical'' distances in the
terminology of \cite{complexA} may over time play a significant role in
astrophysics, just as trigonometric distances did before them, by checking and
calibrating photometric distances. However, before this exciting prospect can
be realised we must overcome the problem that 
the method merely links distances,
velocities and the still uncertain gravitational potential of the Galaxy.
How can we most effectively exploit this link between  measurements of radial
velocities, proper motions and photometry
to obtain tight constraints on both the distances and the potential? This is
an extremely important question given current interest in mapping the Galaxy's dark
matter through the gravitational force field that it generates.
Undoubtedly, pinning down the potential will be greatly facilitated by
modelling several streams simultaneously. 

\item[3.]  Finer discrimination between candidate orbits would be possible if
one could lower the noise floor on the diagnostic function $D'$ by upgrading
the numerical methods used to obtain solutions of equations (\ref{eq:basicde}).
The scheme for searching the space of possible tracks could also be made
faster and more reliable.

\end{itemize}

The list of streams to which this technique could be applied is already quite
long: obvious examples include the
tidal tails of the  globular clusters Palomar 5 and NGC 5466, those of the Large
Magellanic Cloud, the Orphan Stream, and the tidal stream of the
Sagittarius dwarf galaxy. However, before we can exploit the methods
presented here, radial velocities
are needed along these streams.  The greatest precision is promised by the
narrowest streams, and these are well defined only in main-sequence stars.
Therefore the observational challenge is to obtain accurate radial velocities for
tens of faint stars. This will require $8\,$m telescope time for what may
seem unglamorous work. This paper suggests, however, that the scientific rewards of
such observations would be far-reaching. 

Recently it has been realised that
orbits reconstruction is possible using proper motions along the stream
rather than radial velocities \citep{Paper3}. The principles developed in
the present paper will undoubtedly transfer to orbit reconstructions from proper
motions. It remains to be discovered whether they will enable useful
distances to be determined from proper motions of currently attainable
precision. We hope to report on this issue shortly.

\section*{Acknowledgments}

AE acknowledges PPARC/STFC for partially funding this work, Carlo
Nipoti for providing a copy of the FVFPS code, and John Magorrian for
useful discussions about numerical methods. We thank the anonymous
referee for his/her suggestions.

\label{lastpage}

\end{document}